\documentstyle[11pt,aaspp4]{article}

\slugcomment{Submitted to the Astronomical Journal}

\lefthead{Gerardy et al.}
\righthead{Near-Infrared Spectrum of SN 1998S}
\begin{document}

\title{Detection of CO and Dust Emission in Near-Infrared Spectra of 
SN~1998S }
\author{Christopher L. Gerardy \& Robert A. Fesen}
\affil{6127 Wilder Laboratory, Physics \& Astronomy Department \\
       Dartmouth College, Hanover, NH 03755}
\and
\author{Peter H\"{o}flich \& J. Craig Wheeler}
\affil{Department of Astronomy, University of Texas, Austin, TX 78712}

\begin{abstract}

Near-infrared spectra (0.95 -- 2.4 \micron) of the peculiar Type~IIn supernova 
1998S in NGC~3877 from 95 to 355 days after maximum light are presented. 
\textit{K}-band data taken at days 95 and 225 show the presence of the first 
overtone of CO emission near 2.3~\micron, which is gone by day 355. An apparent
extended blue wing on the CO profile in the day 95 spectrum could indicate a 
large CO expansion velocity ($\approx$ 2000 -- 3000 km~s$^{-1}$). This is the 
third detection of infrared CO emission in nearly as many Type~II supernovae 
studied, implying that molecule formation may be fairly common in Type~II 
events, and that the early formation of molecules in SN~1987A may be typical 
rather than exceptional.   Multi-peak hydrogen and helium lines suggest that 
SN~1998S is interacting with a circumstellar disk, and the fading of the red 
side of this profile with time is suggestive of dust formation in the ejecta, 
perhaps induced by CO cooling.  Continuum emission that rises towards longer 
wavelengths (J $\rightarrow$ K) is seen after day 225 with an estimated 
near-infrared luminosity $\gtrsim 10^{40}$ erg~s$^{-1}$.  This may be 
related to the near-infrared excesses seen in a number of other supernovae.  
If this continuum is due to free-free emission, it requires an exceptionally 
shallow density profile. On the other hand, the shape of the continuum is well 
fit by a 1200 $\pm 150$ K blackbody spectrum possibly due to thermal 
emission from dust.  Interestingly, we observe a similar 1200~K blackbody-like,
near-infrared continuum in SN~1997ab, another Type~IIn supernova 
at an even later post-maximum epoch (day 1064+).  A number of dust 
emission scenarios are discussed, and we conclude that the NIR dust continuum 
is likely  powered by the interaction of SN~1998S with the circumstellar
medium. 
\end{abstract}


\section{Introduction}

Detection of molecular emission in SN~1987A opened up a new era in
the study of supernovae (SNe). Analysis of molecular emission in 
SNe can provide a wealth of unique information about the progenitor's interior 
structure, the explosion mechanism and, in general, about the dynamics 
of the explosion. In addition, the presence of molecules can 
directly affect the physical conditions in the SN ejecta.
 
For example, carbon monoxide (CO) can be used as a sensitive probe of 
mixing in the helium cores of supernovae.
The presence of  He$^{+}$ or He$^{++}$ that is
microscopically mixed within the carbon-oxygen layers will destroy 
the CO molecules via charge transfer reactions. Therefore, the presence
of CO establishes that such microscopic mixing has not occurred 
(\cite{lepp90}; \cite{gearhart99}) or, if it has, the helium has not 
been ionized.  In SNe~II, the ionization of He requires non-thermal 
processes, either strong $\gamma$ radiation from the decay of radioactive 
elements such as $^{56}$Ni, or non-thermal electrons. Clumps of nickel 
within a C/O/He mixture could ionize the helium, so the presence of 
CO puts limits on the optical thickness of such clumps and hence the 
amount of mixing of Ni-rich layers.  For non-thermal particles to have an 
effect on CO emission, the inner Ni-rich layer must itself be microscopically 
mixed with both the He and the C/O because of the short stopping length of 
non-thermal particles in the presence of any finite magnetic field. 

Band profile modeling of the CO emission can also be used to infer
the excitation temperature of the molecular emission.  Transitions 
from different vibrational levels create a fine structure in the 
near-infrared bands, and the relative strength of the different band 
heads can be used as a temperature diagnostic (Sharp \& H\"oflich 1989). 
From the width of fine structure CO features, one can obtain information about 
the extent (in velocity space) of the carbon and oxygen rich layers.
In addition, both the total mass and the excitation temperature can be 
inferred from late, optically thin CO emission.  In SN~1987A, \cite{liu92} 
derived a mass $M_{CO} \approx 10^{-3} M_{\sun}$ and a temperature 
$T \approx 4000$ K at $t=192^{\rm d}$ that declined to $T \approx 1800$ K 
by $t=337^{\rm d}$ (but see \cite{gearhart99}).  Similarly, \cite{SL96}
(hereafter SL96) inferred a CO mass of $M_{CO} \approx 10^{-4}
M_{\sun}$ and a temperature of 3000--4000 K, from their spectrum of
SN~1995ad.

Molecules can also act as powerful coolants of supernova ejecta since
they possess a large number of collisionally excitable transitions.  
In fact, in relatively cool ejecta (a few thousand K), CO emission 
could well be the dominant cooling process. This cooling, in turn,
might pave the way to dust formation in SNe. Thus, to compute accurate 
radiative transfer models in the cooling ejecta at late times, it is important 
to include and understand the molecular formation, destruction, and emission 
processes that occur in supernovae (Liu, Dalgarno \& Lepp 1992; SL96; 
\cite{liu94}; \cite{liu95}; \cite{gearhart99}).

Unfortunately, few near-infrared (NIR) spectra of supernovae currently
exist in the literature, making molecule formation in supernovae a
poorly studied phenomenon. To date, molecular emission has only 
been seen in the spectra of two supernovae, both of Type~II.  
Molecules of CO, SiO, and H$_{3}^{+}$ were seen in SN~1987A (\cite{CG87}; 
\cite{spyro88}; \cite{MC93} and references therein) and CO was detected 
in SN~1995ad (SL96).  In this paper, we present evidence for CO emission 
in a third supernova: the Type~IIn event SN~1998S.

SN~1998S was discovered in NGC~3877 by Z. Wan (\cite{Li98};
\cite{qiu98}) on 2~March 1998.  It was classified as a Type~IIn
supernova (\cite{schlegel90}) by \cite{FM98}, but it soon became 
apparent that SN~1998S was a peculiar variant of the Type~IIn class.  
\cite{Garn98a} reported unusually strong \ion{He}{2}~(4686~\AA) and 
permitted \ion{N}{3}/\ion{C}{3}~(4640~\AA) lines in a pre-maximum spectrum, 
as well as broad emission at 5800~\AA \ and 7100~\AA \ due to 
\ion{C}{4} and \ion{C}{2} lines often found in Wolf-Rayet stars.  
Near maximum light, the \ion{He}{2} (4686~\AA) and \ion{N}{3}/\ion{C}{3} 
(4640~\AA) features had nearly disappeared and the H$\alpha$ and H$\beta$
profiles had become asymmetric with steep blue wings (\cite{Cao98}). 
Shortly after maximum, the optical spectrum developed a deep broad
absorption near 6286 \AA \ which was identified as the \ion{Si}{2}
(6347, 6371 \AA) blend often seen in Type~Ia
supernovae (\cite{GJK98}).  Optical and near-infrared spectra taken
near day $\sim$ 230 revealed prominent H$\alpha$ and He I 10830 \AA\
emissions (\cite{G98b}, hereafter G98b) with line profiles unlike
those seen in the UV lines (\cite{GCK98}, hereafter GCK98). G98b report
that the H$\alpha$ emission had a full width at its base of 14~400 km 
s$^{-1}$, with three sharp and well-resolved peaks at --4650, --400, 
and +3700 km~s$^{-1}$, possibly indicating interaction with a circumstellar 
disk or ring.

\section{Observations and Data Reduction}

Low-dispersion optical and near-infrared (NIR) spectra of SN~1998S were
obtained with the MDM Observatory 2.4m Hiltner telescope at Kitt Peak, Arizona.
Table 1 lists the log of spectroscopic observation times and parameters. 
Optical spectra were collected using the Modspec grating spectrograph in two 
low resolution ($\approx 8.5$ \AA ) setups.  The optical data were flux 
calibrated using \cite{Massey88} standard stars and the long wavelength 
extensions of \cite{MassGron}.

Near-infrared observations were made with the OSU-NOAO Infrared
Imager/Spectrometer (ONIS), a high-throughput infrared imager/spectrograph 
with an ALLADIN 512x1024 InSb detector.  This instrument 
can be operated with standard \textit{J}, \textit{H}, and \textit{K} filters 
for broadband imaging, or with a variety of grisms, blocking filters and an 
east-west oriented $1\farcs 2$ and $0\farcs 6$ slits, allowing low 
($R \approx 700$) and moderate ($R \approx 1400$) resolution spectroscopic 
observations from 0.95 to 2.5~\micron.   
 
The near-infrared spectra were corrected for telluric absorption by
observing nearby A~stars, and early G~dwarfs from the Bright Star
Catalog (\cite{BSC}.)  Applying the procedure described by
\cite{HRL98} (hereafter HRL98), stellar features were removed from 
the G~dwarf spectra by dividing by a normalized solar spectrum
(\cite{solar1}; \cite{solar2}\footnote{NSO/Kitt Peak FTS data used
here were produced by NSF/NOAO.}). The resulting spectra were 
used to correct for telluric absorption in the A~stars.  The hydrogen
features in the corrected A~star spectra were removed from the
raw A~star spectra and the results were used to correct the target 
data for telluric absorption. [For further discussion of this procedure 
see HRL98, \cite{HCR96}, and references therein.] The instrumental 
response was calibrated by matching the continuum of the A~star telluric 
standards to the stellar atmosphere models of \cite{kurucz}.

Absolute flux levels for the spectra were set by matching the flux 
to the broadband photometry averaged over a square bandpass.  Using 
ONIS in imaging mode, we obtained \textit{J}, \textit{H} and \textit{K}
images of SN~1998S and HST/2MASS photometric standards (\cite{persson98}) 
yielding photometry concurrent to the spectra.  The photometric results 
are presented in Table~2 and absolute flux levels obtained for the spectra 
are believed accurate to $\sim$ 20--30\%.

The data are arranged into observations of four epochs, based on a date
for $V_{\rm Max}$ of 18 March 1998. (Garnavich, priv. comm.) 
Although observations of the various wavelength regions were sometimes 
spaced a few days apart, within each epoch we treat the data as concurrent 
and refer to it by a single average date. This grouping of the data 
seemed reasonable, as SN~1998S was only observed at late times when
it was evolving relatively slowly and changed little on a timescale of days. 
The specific observation times for the various wavelength regions are listed 
in Table~1.   

While most spectra were obtained in photometric or near
photometric conditions, the day 260 spectrum was observed through
variable cirrus. As a result, these data are quite noisy, and
likely contain residual telluric features. The data from days 95
and 225, while observed under good weather conditions, were taken at
rather high airmass, as SN~1998S was near the Sun in both June and
October. The day 355 spectrum is an average of a large number of 
observations made over a three night period and has the best signal 
to noise of the post 200$^{\rm d}$ spectra.

\section{Results \& Discussion}

Reduced near-infrared and optical spectra of SN~1998S are presented in 
Figures 1 through 4. All have been Doppler corrected for the 902~km~s$^{-1}$ 
redshift of the host galaxy (\cite{DVetal91}). The time evolution of SN~1998S 
in the NIR is shown in Figure 1 with the individual spectra vertically offset 
for clarity and respective flux zero points marked for each plot.
Identifications for prominent emission features in the NIR data for the first 
two epochs are shown in the blow-up of Figures 2 and 3. A list of detected 
near-infrared lines is given in Table~3. All the NIR data are presented in 
vacuum wavelengths.  Figure 4 shows a plot of the day 225 optical spectrum with
line identifications for most emission features marked. 

Our day 95 spectrum (Fig. 2) is in good agreement with the roughly concurrent 
data obtained by Meikle and collaborators (priv. comm.). It also 
shows many similarities to the NIR spectrum of SN~1987A at day 192 
(\cite{meikle89}, hereafter M89).  In fact, virtually all the features 
seen in this spectrum can be identified in the day 192 spectrum of SN~1987A
(M89.)  Strong features include Pa$\beta$ (1.282~\micron), 
\ion{Mg}{1} (1.183~\micron\ and 1.503~\micron), Br$\gamma$ 
(2.166~\micron), and the first overtone of CO, which rises
near  2.28~\micron. The 1.183~\micron\ \ion{Mg}{1} feature appears 
shifted somewhat to the red which may be due to blending with
\ion{Si}{1} lines near 1.2~\micron.  This feature was also observed shifted 
to the red in the SN~1987A spectrum (M89.) The bright feature seen
coming up at the blue end of the \textit{J}-band spectrum near
1.13~\micron\ is likely \ion{O}{1} (1.1287~\micron) and the faint
emission at 2.06~\micron\ is probably \ion{He}{1} 2.056~\micron.  
The broad emission in the \textit{H}-band, which rises at $\sim
1.57$~\micron, is unlikely the second overtone CO emission  
(\cite{gerardy98}), but rather the blend of \ion{Si}{1},
[\ion{Si}{1}], [\ion{Ca}{2}], [\ion{Fe}{2}], \ion{Mg}{1}, and 
Brackett series lines seen in the SN~1987A spectrum of M89. In SN~1987A, 
this feature was much less smooth, but differences with SN~1998S 
can be explained if the Brackett lines are weak and [\ion{Fe}{2}] 
emission is weak or absent. This is consistent with Br$\gamma$ being 
significantly weaker in SN~1998S than in SN~1987A. 

Evolution of SN~1998S in the near-infrared between days 95 and 355 
is apparent from Figure 1.  Many of the broad emission lines seen in the day 
95 data are absent in later spectra, with the spectrum longward of 
1.3~\micron\ dominated by continuum emission.  At day 225, the NIR continuum 
rises from $\lambda\approx 1~$\micron\ to $\lambda\approx 1.75~$\micron, where 
it then levels out to a plateau in the \textit{K}-band.  Although at day 
260 the rise to the red appears much weaker in the 1--1.8~\micron\ region,
this is likely an artifact of the poor observing conditions experienced when 
these data were obtained. The continuum level in the \textit{K}-band measured 
via photometry indicates that the flux continues to rise from \textit{J} 
through \textit{H} to \textit{K}.  The late-time shape of the NIR continuum of 
SN~1998S is most clearly seen in the higher signal-to-noise spectrum at day 
355.  Here the rise in the continuum appears almost linear, with a 
somewhat shallower slope in the \textit{K}-band.

While the continuum is the dominant NIR feature in the later data, 
a few line emission features remain. The first overtone CO emission although 
fainter, is still present in the day 225 spectrum, as is the 1.13~\micron\ 
and 1.19~\micron\ emission seen in the 95$^{\rm d}$ spectrum.  
No trace of these features is seen in the 260$^{\rm d}$ and 355$^{\rm d}$
data. The most prominent emission lines in the post 200$^{\rm d}$ spectra
are two features with multiple peaks near 1.28~\micron\ and 1.08~\micron.
The first is Pa$\beta$ and the second is a blend of \ion{He}{1}
(1.083~\micron) and Pa$\gamma$.

A multi-peak line profile can also be seen in the optical spectrum on day 225 
(compare Fig. 3 and Fig. 4).  H$\alpha$, the dominant emission feature,
exhibits a line profile with three strong peaks at --4900 km~s$^{-1}$,
--300 km~s$^{-1}$, and +3300 km~s$^{-1}$, in rough agreement with those 
reported by G98b.  Comparison of our data with that described by G98b
indicate that measured velocity differences are due to real changes in the 
line profile over the 17 days between the two data sets 
(Garnavich, priv. comm.). This three-peak profile is also seen in H$\beta$ 
and \ion{He}{1} (5876 \AA).  In the 225$^{\rm d}$ NIR spectrum (Fig. 3), the 
1.08~\micron\ feature has four peaks.  The first three of these peaks are 
\ion{He}{1} 1.083~\micron\ emission peaked at --4900, --500, and 
+3700 km~s$^{-1}$, matching the emission profile of H$\alpha$ to within our 
measurement errors.  The longest wavelength peak of this 1.08~\micron\ feature 
corresponds to Pa$\gamma$ emission at +3900~km~s$^{-1}$ and is probably the red
edge of a three-peaked Pa$\gamma$ profile, with the other two peaks being lost 
in the \ion{He}{1} 1.083 \micron\ emission.  The Pa$\beta$ profile could be 
similarly triple peaked, but the line is in a region of fairly strong telluric 
absorption which is difficult to accurately remove from such a faint target.  
The strong peak at the red end of this feature corresponds to a Pa$\beta$ 
velocity of +3200~km s$^{-1}$, which is close to the velocity of the 
red peak in the H$\alpha$ profile. 

Although the optical emission is dominated by bright H$\alpha$
emission, other features can be seen (Fig 2.) These include broad emission
from [\ion{O}{1}] (6300, 6364 \AA), [\ion{Ca}{2}] (7291, 7334 \AA), and 
\ion{Ca}{2} (3933, 3969 \AA).  [\ion{O}{3}] (4969, 5007 \AA) is present as 
narrow emission, and possibly has a broad component as well, but the feature 
is blended with H$\beta$.  Two very broad emission features from 
$\approx$ 4200 to 4700 \AA\ and from $\approx$ 5100 to 5400 \AA\ are a blend 
of [\ion{Fe}{2}] emission as reported by GKC98.  Finally, we attribute the 
very broad feature from $\approx$ 8250 to 8800 \AA\ to a blend of emission 
from \ion{O}{1} 8446 \AA, [\ion{C}{1}] 8727 \AA, and the \ion{Ca}{2}
IR triplet 8498, 8542, \& 8662 \AA.

\subsection{Carbon Monoxide Emission}

A full quantitative analysis of the observed CO emission requires detailed 
models for the explosion including any asphericity in the envelope, the spatial
and time variation of the chemical structure and, in the case of
SN~1998S, a good model for the interaction with the environment
(see \S 3.2 below).  As these details are not well constrained in SN~1998S, 
we will restrict ourselves to a fairly basic analysis.  From the presence of CO 
emission, we know that ionized helium was not mixed with the underlying C/O rich
layers on a microscopic scale (\cite{lepp90}).  Investigations of
SN~1987A (\cite{HSZ89}) concluded that solar C and O abundance could not produce
CO features as strong as observed.  This suggested that the abundances
of C and O must be enriched, and that CO formation was being observed within the
C/O-rich layers of the ejecta of SN~1987A.  For SN1998S, neither the density 
slope nor the mass above the photosphere are well determined. 
Consequently, alternative explanations for the strength of the CO feature may be
valid.  In particular, a larger mass above the photosphere or a steeper density 
slopes in comparison to SN1987A would increase the C/O density.  

Portions of the \textit{K}-band spectra from days 95, 225, and 355 are
displayed in Figure 5, showing the evolution of the CO emission.
In the 95$^{\rm d}$ spectrum, the CO feature declines to the red and 
is similar to the 192$^{\rm d}$ spectrum of SN~1987A (M89); however, roughly
concurrent data obtained by Meikle's group show a flatter CO profile 
(priv. comm.). We have some concern that the red end of our CO data 
may be contaminated by residual telluric absorption.  In the day 225 data, the 
CO feature is much fainter, and consequently the signal-to-noise is worse. 
Although the CO emission profile at day 225 looks relatively flat, the data 
are poorly constrained at the red end and thus we cannot be certain about its 
true shape.  A flat emission profile would be in sharp contrast to the 
evolution of SN~1987A where the red end faded significantly before the blue 
(M89).  Finally, by day 355 the CO feature faded below our detection limit.  
It is possible that the CO feature faded much earlier since there is no clear 
evidence of it in the day 260 spectrum, but these data are too noisy to rule 
out the presence of CO emission.

\subsubsection{CO Temperature}
The shape of the CO profile can constrain the temperature of the CO emitting
region.  LTE emission modeling of CO by Sharp and H\"oflich (1989) and 
\cite{liu92} has shown that if the temperature is much below 2000 K, the first
two vibrational bands (v = 2 -- 0, 2.2935~\micron\ and v = 3 -- 1, 
2.3227~\micron) have comparable flux levels while vibrational bands 
three and higher are significantly weaker by at least 20 - 30 percent. This
leads to a noticeable break in the emission profile redward of the second band 
head.  The 95 day spectrum of SN~1998S shows no sign of such a 
break.  This flux distribution is consistent with a lower limit for the 
temperature at day 95 of $\approx 2000$~K assuming LTE in the vibrational bands
(\cite{SH89}).  If the flux distribution longward of the band head is flatter,
as Meikle's data suggest, a higher temperature is indicated.   

The CO emission profile for day 225 is too noisy and uncertain to ascribe any
CO temperature. We are therefore unable to determine if the temperature of the
CO emitting region has changed, as it had for SN~1987A from about 3200 K to
1700 K during the interval between day 192 and 283
({Liu, Dalgarno \& Lepp 1992}).

\subsubsection{Comparison to SN~1987A and the CO velocity}
With so few observations of CO emission in supernovae, it is useful to compare 
the observed CO emission in SN~1998S to that seen in SN~1987A. In Figure 6 we 
compare the observed CO spectrum of SN~1998S at day 95 to the day 192 CO 
spectrum of SN~1987A presented by M89.  We have shifted both spectra to the 
supernova rest-frame, and we have scaled the
SN~1987A data to match the SN~1998S data at the continuum near 2.24 \micron \ 
and near the CO peak in the 2.30 -- 2.32 \micron \ region. This plot shows that
longward of the steep rise near 2.28~\micron \ the profiles seem 
(at least qualitatively) quite similar, although our spectrum of SN~1998S 
does not show convincing evidence for strong CO band structure like that seen 
in SN~1987A.  The narrow minima in the SN~1998S spectrum do not correspond to 
physical minima between bands and hence are most likely due to noise and 
residual telluric absorption features.   Model calculations show that 
peak-to-trough variations at the level exhibited by SN~1987A result from a 
CO expansion velocity $\approx 2000$ km~s$^{-1}$.  It is possible, 
however, that the noise in our SN~1998S data could obscure the bandhead 
structure at this level.  Much lower CO velocities, like the $\approx 500$ 
km~s$^{-1}$ CO velocity seen in Nova Cas 1993 for instance, can certainly be 
ruled out as the bandhead structure would be quite distinct (\cite{evans96}).

Shortward of the rise near 2.28 \micron, the SN~1987A and SN~1998S spectra 
in Figure 6 show significant differences.  Although both spectra show excess 
emission in the 2.25 -- 2.28 \micron \ region, in SN~1987A this emission is 
due to an unidentified emission line which peaks around 2.65 \micron, while in 
SN~1998S the emission in the region consists of a relatively smooth rise to 
the blue edge of the CO feature.  This emission is not explained by 
moderate ($\approx 2000$ km~s$^{-1}$) velocity models of CO emission which 
predict quite a sharp drop to zero at the blue edge of the CO profile (Liu, 
Dalgarno, \& Lepp 1992; \cite{liu95}).  While it is possible that this emission
in SN~1998S is due to noise or the blending of some unidentified feature with 
the CO emission, it certainly does not match the unidentified feature in 
SN~1987A. 

On the other hand, the blue edge of the SN~1998S CO profile appears 
consistent with emission from CO at high velocity 
($\approx$ 2000 -- 3000 km~s$^{-1}$).  To illustrate this point, we have
calculated a grid of CO emission models, varying the velocity, temperature, 
density profile and envelope mass, and compared the results to the observed 
SN~1998S CO emission profile. The details of the model calculations are 
discussed in Appendix A. Some examples of our results are shown in Figure 7 
where the qualitative effects of varying the model parameters can be seen.  Low
temperatures show stronger emission in the lower vibrational bands compared to 
high temperatures and steeper density structures tend to result in a flatter 
decline of the CO-flux towards the red.  Low expansion velocities
preserve the strong band structure of the CO whereas velocities in excess of 
3000 km~s$^{-1}$ almost smear the features out and, in the optically thick
case, result in an offset of the CO features due to the Doppler shift 
corresponding to the photospheric expansion velocity.  Also, as the expansion 
velocity increases, the short-wavelength edge of the profile grows an extended 
blue wing.   

\subsubsection{Comparison to CO models}
Comparisons of several models with the observed day 95 CO profile are shown in
Figure 8.  These models have been taken from our grid of calculations without 
fine tuning the parameters to best match the data.  In most respects, our data 
does not provide strong constraints on the emission model.   Models with a wide 
range of parameters provide equally good fits to the data.  Despite these 
uncertainties, however, we find that steeper density profiles, coupled with 
high expansion velocities tend to give better agreement with the observations.  
In particular, the observed emission at the blue edge of the CO profile, if 
due to CO emission, strongly favors high expansion velocities in the range of 
2000 -- 3000 km~s$^{-1}$.  

We wish to emphasize that the comparison with the observations should not be 
regarded as a true model fit but rather as a proof of principle, as there are a
number of complications that we have ignored in our models.  First, high 
polarization has been observed in SN1998S both by the groups in Berkeley 
(\cite{Leonard99}, hereafter L99) and Texas (\cite{wang99}) which indicate 
deviations from sphericity of the order of 1/2. Second, the density structure 
may be quite clumpy, which whould change the formation time scales for CO as 
several processes go with the square or cube of the density.  In addition, CO 
formation is a strong coolant and may significantly effect the temperature 
structure as in the case of SN1987A (e.g., \cite{hof88}).  A more realistic 
treatment of the CO-spectrum requires full 3-D, NLTE radiation transport 
calculations for the time evolution, which is beyond the scope of this paper.

\subsubsection{Implications of high velocity CO}
Several important implications result if a high velocity CO interpretation is 
correct.  The presence of high velocity CO is inconsistent with a progenitor of
only moderate mass, 10 to 20 $M_{\sun}$, as stars in this mass range have 
relatively little C and O compared to He and H and it is buried deeply within 
any ejecta.  For 10 $M_{\sun}$ of ejecta and $10^{51}$ erg of kinetic energy, 
the mean velocity is about 3000 km s$^{-1}$.  (Note: The mean velocity
typically corresponds to the local velocity about 30 percent of the way out in 
the ejecta mass.) For a star with a main sequence mass of 
$M \lesssim 20$ $M_{\sun}$, the C and O layers are at a mass fraction of 
$\lesssim$ 0.3 (\cite{WW95}), making it difficult to produce substantial CO at 
$\approx 3000$ km s$^{-1}$.   

For higher mass stars, on the other hand, the mass of C and O rises to become 
comparable to that of the helium, and the C/O layers extend comfortably beyond 
the region representing the mean velocity if a substantial amount of the mass 
of the hydrogen envelope has been expelled.  Thus high velocity CO would imply 
that the progenitor of SN~1998S was a substantially massive star, in excess of 
25 $M_{\sun}$, and that the progenitor had lost a substantial portion of its 
hydrogen mass.  This might be consistent with the weakness of the $Br_\gamma$ 
line seen in SN~1998S as compared to that observed in SN1987A, indicating a 
rather small hydrogen abundance.  Furthermore, the hydrogen observed in SN1998S 
may be dominated by emission from the interaction region with the CSM.  This 
might suggest that SN~1998S was a Wolf-Rayet star.  Such an interpretation would
be consistent with the Wolf-Rayet features seen in the early optical spectra 
(\ion{C}{2}, \ion{C}{3}, \ion{C}{4}, \ion{N}{3}) (Garnavich et al 1998a) and is 
also supported by the model of \cite{wang99}.   
 
\subsubsection{Is CO formation common?}
Finally, we note that of the four Type~II SNe for which near-infrared 
spectra have been published,  (SN~1987A, SN~1995ad (SL96), SN~1998S (this
paper) and SN~1995V (\cite{fassia})) this is the third with detected CO 
emission.  While this is certainly far from a statistical sample, it does 
suggest that CO emission may be a fairly common feature in the late-time 
spectra of Type~II SNe.  Obviously, more NIR spectra of Type~II SNe in the 
100 -- 300 day time period are needed before one can determine the generality of
CO emission in these events.

\subsection{Multi-Peak Emission}
Multi-peak emission profiles in the spectrum of SN~1998S are only seen in the 
hydrogen and helium lines.  The [\ion{Ca}{2}], [\ion{O}{1}], and \ion{Ca}{2} 
lines all have profiles that are peaked in the blue and fade to the red.
The fact that we see evidence for the three-peaked profile in the
hydrogen and helium lines, and only in these lines, suggests
it is emission coming from circumstellar gas and not ejecta.
Late-time optical spectra of Type~IIn supernovae are dominated by
emission from the circumstellar interaction (\cite{Fil}). In this
respect, SN~1998S is not an exception, although this three-peaked profile is
certainly a peculiar variation. 

The three peak line profile seen in SN~1998S can be naturally understood as the
result of a supernova shock wave interacting with a disk shaped circumstellar
medium.  The high positive and negative velocity peaks of the hydrogen and 
helium emission line profiles can be explained as emission from an equatorial 
ring.  For a uniformly emitting ring or disk expanding at a constant velocity,
\begin{equation}
 \frac{dL}{d\nu} \propto \left( 1 - \left(
\frac{\Delta\nu}{\Delta\nu_{max}}\right)^{2} \right)^{-1/2},
\end{equation}
(McCray, priv. comm.).
In Fig 9a, we show a plot of this profile.  In reality, the emission from 
shocked gas would have a finite width due to post-shock velocity dispersion. 
This will tend to round off the sharp edges, giving the profile a shape better 
matching the observed H$\alpha$ profile.  If the emission is coming from a thin
annulus with a slight velocity gradient between the outer and inner edges, this
will also tend to soften the peaks and give a somewhat smoother decline to zero
intensity at the very highest velocities. 

The similarity between Figure 9a and the multi-peak line profile
suggests that the circumstellar emission in SN~1998S has a ring-shaped
component.  Such ring-shaped emission could be the natural result of the 
supernova shock wave colliding with a circumstellar ring or disk. While the 
central peak is not explained by such a simple picture, it may be 
related to the moderate-width emission which was the dominant 
late-time feature in the spectrum of the Type~IIn SN~1988Z  (\cite{turatto93}; 
\cite{aretxaga99} and references therein).  The low velocity peak can be 
plausibly explained by the introduction of a multi-component circumstellar 
medium, similar to the ``clumpy wind'' model proposed by Chugai \& Danziger 
(1994) to explain the SN~1988Z emission profile.  

Here we consider the case that the circumstellar disk around SN~1998S had a 
population of dense clouds embedded in a less dense inter-cloud medium.  The 
expanding shock accelerates the intercloud gas to high velocity, which then 
radiates in a thin ring behind the shock front creating the high velocity peaks 
as described above, while the dense clouds undergo significantly less 
acceleration, and emission from these clouds could form the central peak.  

For the shocked clouds, the post-shock velocity
will be related to the cloud density by $ \rho_{c} \propto u_{c}^{-2}$, where
the constant of proportionality is a function of the dynamical 
pressure of the shock.  If we assume a population of clouds that 
obeys a power law number density $n \propto \rho_{c}^{-k} \propto u_{c}^{2k}$ 
with a cutoff at $\rho_{c_{min}} \propto u_{c_{max}}^{-2}$ which we treat as a 
free parameter, then taking the cloud emissivity 
$j \propto \rho_{c}^{2} \propto u_{c}^{-4}$ gives a total flux 
$F_{c}\left( u \right) \propto n\left( u \right) j\left( u \right) \propto 
u^{2k-4}$ from clouds moving at post-shock velocity $u$.  Since the dense 
clouds will also be emitting from a ring or 
disk in the cooling region behind the shock front, the observed line profile is
obtained by convolving $F_{c}\left( u \right)$ with the ring emission profile 
(Eq 1), resulting in  
\begin{equation}
\frac{dL}{d\nu} \left( \Delta\nu \right) \propto \int_{u_{min} 
}^{u_{c_{max}}} \frac{u^{2k-3}}{\sqrt{u^{2} - \left( u_{min} \right) ^{2}}} du,
\end{equation}
where 
$$
u_{min}\left( \Delta\nu \right) = \left( \frac{c \Delta\nu }{\nu_{0}} 
\right). 
$$  

In Figure 9b we overplot a version of this simple model on the observed 
H$\alpha$ profile.  The outer parts of the model are emission from a thin 
annulus with a small (3\%) velocity gradient, and a maximum velocity of 4300 
km~s$^{-1}$.  The central peak is the profile described by Equation 2, with 
$k = 1.9$ and a cutoff velocity $u_{c_{max}} = 1000$ km~s$^{-1}$.  The high 
and low velocity components have been blueshifted by 700 and 400 km~s$^{-1}$ 
respectively to better match the observed profile.  This blueshift might be 
due to either obscuration by dust (see below) or asymmetries in the 
circumstellar interaction. 

The evolution of the triple-peaked circumstellar emission feature is suggestive
of dust formation in the supernova ejecta. In the 260$^{\rm d}$ and 
355$^{\rm d}$ spectra, the red side of the 1.08~\micron\ feature steadily
fades, while the blue peak remains largely unchanged.  Similar evolution was 
seen in the H$\alpha$ and H$\beta$ profiles. (see Garnavich et al. (1999) in 
prep, and L99)  A simple explanation for the evolution of the triple-peak 
profile is that dust has formed in the ejecta, inside the ring shaped emission, 
although the weakening of the red side due to asymmetries in the circumstellar 
interaction cannot be ruled out.  There is relatively little difference between 
the late (red-side obscured) profiles of H$\alpha$ and \ion{He}{1} 1.08~\micron,
suggesting that the dust has formed in optically thick clumps.  A schematic of 
the geometry implied by this picture is given in Figure 10. 
	
Optical spectra taken earlier than our day 225 data may indicate that the red
peak of the H$\alpha$ profile was originally  brighter than the blue peak, 
and then faded (L99).  While this could again be due to asymmetries in the 
circumstellar interaction, it could also indicate the presence of dust in the 
circumstellar disk surrounding SN~1998S.  If the expanding shock either sweeps 
out or destroys the circumstellar dust, and if the line of sight to the 
supernova is close to the plane of the circumstellar disk, then before the 
appearance of the dust in the ejecta, the blue peak would be suppressed
relative to the red peak as observed by L99.    
 
\subsection {Near-Infrared Continuum}
At t $\geq 225^{\rm d}$ both the spectroscopic and photometric data show 
that SN~1998S exhibits a near-infrared spectral energy distribution which 
becomes brighter toward longer wavelengths.   Similar near-infrared excesses 
have been previously observed in the photometry of a number of other 
supernovae including SN~1979C (\cite{Merrill80}), SN~1980K (\cite{Telesco81}), 
SN~1982E (\cite{Graham83}; \cite{GM86}), SN~1982L, SN~1982R (\cite{GM86}), 
SN~1985L (\cite{Elias86}), SN~1993J (\cite{Lewis94}), and SN~1994Y 
(\cite{Garnavich96}). This NIR excess has usually be interpreted as thermal 
emission from dust, either forming in the ejecta, (\cite{Merrill80}; 
\cite{Telesco81}; \cite{Dweketal83}; \cite{Elias86}), or lying in a 
pre-existing circumstellar medium and being heated by the initial supernova 
flash as an infrared echo (\cite{BE80}; \cite{Graham83}; \cite{Dwek83}; 
\cite{GM86}; \cite{Lewis94}; \cite{Garnavich96})  To our knowledge, this is 
the first time a near-infrared spectrum of this phenomenon has been published 
and conclusively demonstrates that this NIR excess, at least in SN~1998S, is 
indeed due to a rising continuum with little contribution from line emission.

\subsubsection {Free-Free Emission}
We begin by considering the plausible emission processes that could explain
the NIR continuum seen in SN~1998S.  At day 355, the rise in the NIR continuum 
flux, $F_\lambda$, with increasing wavelength may be accounted for either by 
free-free emission or by a thermal component due to dust emission. Bound-bound 
and free-free processes are the dominant opacity source in the IR at 
early times. In the low temperature and density of the late time environment 
the bound-free component becomes small in the infrared because the excited 
levels become depopulated.  Free-free emission then dominates at late times in 
the absence of dust. 

In the optically thin limit, continuum flux is proportional to the 
emissivity, $\eta = \kappa\rho S$, where $\kappa$ is the opacity and S the 
source function.  For free-free processes, $\kappa_{ff} \propto \lambda ^2$, 
and the corresponding source function is given by a black body, 
$B_{\lambda} \propto \lambda^{-4}$, because the free electrons are thermalized.
The latter relation holds for the IR if the temperature in the 
continuum-forming region is greater than 3000 K. For this case, 
$F_\lambda \propto \eta \propto \lambda ^{-2}$. Thus, we can exclude this case 
for all the later spectra and especially for day 355.
 
If the continuum is optically thick in the IR due to free-free radiation, 
the flux is $F_\lambda \propto R(\lambda )^2 B_{\lambda}$, where 
$R(\lambda)$ is the wavelength-dependent photospheric radius.  The flux 
then depends on the radial depth from which the emission arises.  For 
free-free absorption, $\kappa_{ff} \propto T^{-3/2}n_e$.  We assume 
that $T(R) \propto R^{-m}$ where $m = 0$ would correspond to constant 
temperature as might represent a strongly NLTE atmosphere and $m = 1/2$ 
would correspond to an LTE atmosphere with constant luminosity. Finally, we
assume a power law density profile $\rho \propto R^{-n}$.  

There are two limiting cases we can consider for 
the electron density, one in which the electron density is constant, due, 
for instance, to ionization by $\gamma$-rays, and one in which the electron 
density scales with the total (ion) density.  The corresponding flux ratio at
two wavelengths of the same optical depth ($\approx$ 1) is then:    
\begin{equation} 
F_{\lambda_1}/F_{\lambda_2} = \left( \lambda_1 \over 
\lambda_2 \right)^{16-8nk + 12m \over 2nk - 3m -2}, 
\end{equation} 
where $k = 1$ corresponds to the case of $n_e =$ constant and $k = 2$ 
corresponds to the case with $n_e \propto \rho$.  This expression gives 
a constant slope for $n \lesssim 2$.  

For the spectra of day 355, $F(2.4~\mu m) \approx 4.1 \times F(1.2~\mu m)$. 
We find that if the rise in the continuum to the red in the day 355 spectrum 
is due to optically-thick, free-free emission, the density slope must be 
fairly flat, $n \lesssim$ 1 to 2.  The only exception for this simple model 
is the case where $k = 1$ (constant electron density) and $m = 1/2$
(LTE atmosphere with declining T), for which $n$ could be as large as 2.5, 
still a rather shallow density gradient. Thus, if the continuum at day 355 is 
due to optically-thick free-free emission, then as the ejecta expand and turn
optically thin, the continuum slope must decrease. 
If such a decline is not seen, then the implication is that the continuum 
is not due to free-free radiation.

At this time, optically thick free-free emission cannot be ruled out, 
although the required flat density slopes at the NIR
photosphere are rather extreme.  It is possible 
that the ejecta could remain optically thick at day 355, but to check this 
quantitatively requires a better understanding of the inner density and
velocity profiles than are available.  Any such estimates are also complicated 
by possible asymmetries in the ejecta.

\subsubsection{Thermal Dust Emission}
An alternative explanation for the late-time NIR continuum is thermal dust 
emission. If the continuum at day 355 is due to thermal emission from 
a dust of carbonates, it must have a corresponding blackbody temperature 
less than the 1600 K evaporation temperature (\cite{gail84}). The peak 
of such a blackbody is at about 1.8 \micron. Since we see no obvious continuum
peak at this or shorter wavelengths, any dust in SN~1998S must be cooler than 
1600 K. Carbonates start to condense below 1400 to 1100 K. T=1400 K corresponds
to a Wien peak at 2.07 \micron, so even this temperature may be an upper limit 
for our data.  

We tried matching the day 355 data to blackbody curves with temperatures 
ranging from 1000 to 1400 K.  This was complicated somewhat by the relatively 
large uncertainty in the relative flux level between the \textit{K}-band and 
the shorter wavelength 0.95--1.75 $\micron$  region. Within a 
wavelength region that is observed with a single spectroscopic setting, the 
relative flux is set by the instrumental response and the telluric absorption 
correction, and should be accurate to $\sim$5\% or better (at least outside of 
the regions of strong telluric absorption). Between data taken with 
different spectroscopic settings, however, the relative flux is only set by the
photometric flux calibration, and is therefore subject to the 20--30\% 
uncertainty in the absolute flux levels.  Thus between the 
0.95--1.75~\micron\ and 1.99--2.40~\micron\ regions (see Table 1) the relative 
fluxing is rather uncertain, and in matching the data to blackbody curves we 
treated the absolute flux level of the \textit{K}-Band as a free parameter as 
long as the shifts in the data were less than about 30\%.  So for each 
temperature, the flux level was matched to the \textit{H}-band region, and the 
\textit{K}-band was then lowered or raised to match the level of the blackbody 
curve.  For each temperature we could then judge the `fit' by how well the 
blackbody curve matched the shape of the resulting (flux level adjusted) 
spectrum.

This procedure revealed that the 355$^{\rm d}$ NIR continuum seen in SN~1998S 
could be well matched by a single blackbody with a temperature of 
$1200 \pm 150$ K.  This is illustrated in the top half of Figure 11, where a 
1200 K blackbody curve is overlayed on first the unshifted data (top spectrum),
and then on the same spectrum with the \textit{K}-band region shifted down 16\%
(second spectrum from the top). Therefore the rise in the continuum towards the
red in the day 355 data is consistent with, but does not prove the existence
of, emission from hot dust. 

Figure 11 also shows the late-time NIR spectrum for another Type IIn supernova,
SN~1997ab.  SN~1997ab was discovered on 11 April 1996 in objective prism 
data taken for the Hamburg Quasar survey, and was classified as a 
``Seyfert-1'' Type II supernova (Type~IIn) after a spectroscopic follow up 
on 28 February 1997 (\cite{HR97}). Because it was quite luminous, 
$M_{B} = -19.1$ on 11~April~1996 and faded to only $M_{B}=-17.6$ by 
2~March~1997 it was linked with the slow-fading SN~IIn like SN~1988Z and
SN~1987F (\cite{HER97}).  

Our NIR spectra of SN~1997ab were taken on the same nights as those of
SN 1998S at day 355, and reduced in the same fashion. One sees a very similar 
NIR continuum in SN~1997ab even though it was observed at a substantially later
time (day 1064+).  In fact, applying the same blackbody matching procedure to 
SN~1997ab as we did to SN~1998S again yielded a 1200 K temperature, although
the temperature is probably a bit less constrained due to the lower 
signal-to-noise of the SN~1997ab spectrum. The striking similarity of the NIR 
continuum of SN~1998S and that of SN~1997ab suggests that the continuum in both 
objects is caused by the same physical process.  
 
Assuming the NIR continuua of these objects are really due to 1200 K blackbody
emission, then the implied NIR luminosities are $4.5 \times 10^{40} h^{-2}$ 
erg~s$^{-1}$ for SN~1998S, and $7.9 \times 10^{41} h^{-2}$ erg~s$^{-1}$ for
SN~1997ab. If the assumption of a blackbody is wrong, a lower limit on the
NIR continuum luminosities can be estimated from just the observed 
0.95--2.40~\micron \ flux, yielding luminosities of $1.1 \times 10^{40} h^{-2}$
erg~s$^{-1}$ for SN~1998S and $2.0 \times 10^{41} h^{-2}$ erg~s$^{-1}$ for 
SN~1997ab. In either case, the NIR luminosities for these two SNe are 
considerable.

Because SN~1998S and SN~1997ab have large NIR luminosities at such 
late time, we can rule out radioactive decay as the power source for the 
near-infrared continuum.  This can be seen through a comparison of our observed
NIR luminosities to the total radioactive power of SN~1987A at epochs similar 
to the ages of SN~1998S and SN~1997ab.  For
$18^{\rm d} \lesssim t \lesssim 1200^{\rm d}$, the dominant energy source in
SN~1987A was $^{56}$Co decay, and the total radioactive power was
$\approx 5 \times 10^{40}$ erg~s$^{-1}$ at t~$\approx 350^{\rm d}$, which
declined to $\approx 10^{38}$ erg~s$^{-1}$ by t~$\approx 1100^{\rm d}$
(\cite{MC93} and references therein).  While this is just barely be enough
to explain the SN~1998S luminosity at day 355 if all of the radioactive
energy went directly into the NIR continuum via dust heating, it fails to
explain the SN~1997ab luminosity at t~$\gtrsim 1064$ by five orders of
magnitude. Obviously a different power source is needed to explain this late
time continuum.  

\subsubsection{Dust Emission Scenarios}
If the continuum is due to thermal emission from dust, then one can ask the
following: Where is the dust is located, how is it being heated, 
and why is the dust temperature the same in two objects that are of 
considerably different age?  Here we discuss possible scenarios that could 
explain this late-time NIR continuum.  

Perhaps the simplest model is that we are simply seeing emission from newly
formed dust in the ejecta.  Many novae display similar infrared excesses, often
accompanied with a coincident drop in the optical flux, and this has been
interpreted as evidence for dust forming in the expanding nova shells 
(\cite{G88}, hereafter G88).  This newly formed dust almost always has a 
temperature of T $\approx$ 1000--1200 K, suggesting that this is the 
characteristic temperature at which conditions are right for dust to condense
(G88).  In the 1--2.4 \micron \ region, thermal emission would be dominated by
this hottest component of the newly formed dust, and so the similar
temperatures
seen in both SN~1998S and SN~1997ab might be interpreted as an indication that
dust is still forming in both of these objects.  However, such an
interpretation
would require that either the epoch of dust formation is quite long, (on the 
order of years) or that we coincidentally observed both SN~1998S and SN~1997ab
just as the dust was forming.   Furthermore, this scenario still leaves the 
question of the very large late-time energy source for SN~1997ab unanswered.
 
Another possibility is that the emission is an infrared echo.  
In this scenario, dust formed in the ejected circumstellar 
envelope long before the explosion, and it is heated by the 
initial UV/optical flash of the supernova as it propagates outward 
into the surrounding material (\cite{BE80}; \cite{Dwek83}). The late-time
energy source is then the luminosity of the supernova near maximum
light, perhaps solving some of the energy requirement problems.  
For a spherically symmetric distribution of dust, the IR echo is dominated
by emission from the hottest dust, and as the IR echo ages, the observed 
temperature should drop due to geometric dilution (\cite{GM86}).  This model 
has been successfully used to explain the NIR light curve of SN~1982E 
(\cite{Graham83}; \cite{GM86}).  This interpretation,
however, would again require that the identical temperatures seen in both 
SN~1998S and SN~1997ab are purely accidental. Also, it is debatable
whether the large observed NIR luminosity of SN~1997ab more than three years 
after maximum can be explained in the infrared echo model. 

A third scenario is suggested by the Type~IIn classification of SN~1998S and 
SN~1997ab.  As both events show evidence of strong interaction with a dense
circumstellar medium, it is plausible that the NIR continuum is due to dust
heated by the circumstellar interaction.  This interaction could power the
observed NIR emission either directly through shock heating, or by heating the
dust with X-ray emission from the interaction region.  If the dust is X-ray 
heated, then we would expect an X-ray luminosity at least as large as the 
observed NIR luminosity. The large observed NIR fluxes for SN~1998S and 
SN~1997ab might be consistent with the late-time X-ray luminosities seen in 
some Type IIn objects. The X-ray luminosity SN~1988Z, for example, was
$\sim 10^{41}$ erg~s$^{-1}$ more than six years after maximum (\cite{FT96}),
so the large NIR luminosity of SN~1997ab three years after maximum could be
powered by a similarly strong interaction.  While we can not rule out X-ray 
heating, the direct conversion of bulk flow kinetic energy to heat through
shock heating is certainly a more energetically economical scenario.  

In either case, the dust could be pre-existing dust in the CSM, or newly formed
dust in the supernova ejecta.  Circumstellar dust could be heated by the outer 
blast wave, while dusty ejecta could be heated by a reverse shock traveling 
backwards into the supernova envelope.  Likewise X-rays could heat any dust 
with a relatively unobscured line of sight to the interaction region, and so 
the NIR emission could be from either a dusty CSM or dusty ejecta.  As
discussed above (\S 3.2) the decline of the red-peak of the three-peak profile 
is strong evidence for the formation of dust in the ejecta of SN~1998S. While 
this dust could certainly be the source of the NIR emission, we cannot rule out 
the possibility that the emitting dust was pre-existing dust in the 
circumstellar disk.  L99 inferred quite high densities 
($n_{e} \approx 10^{7}$ cm$^{-3}$) for the CSM, and it would not be unreasonable
to suppose that dust formed in such an environment.  Also, as discussed above, 
asymmetries in the H$\alpha$ seen by L99 in spectra that pre-dates our day 225 
data might be indicative of dust in the circumstellar disk around SN~1998S.    

In principle, since we know both the temperature and the luminosity, we can 
infer an effective surface area for the emitting region. If the emitting region
is a sphere, then the luminosity is given by 
$L = 4 \pi R_{eff}^{2} \sigma T^{4}$, which yields effective radii of 
$370 h^{-1}$ A.U. for SN~1998S, and $1500 h^{-1}$ A.U. for SN~1997ab. 
Both are consistent with ejecta expanding at an average velocity of
$\sim 2000$ km~s$^{-1}$ since the explosion.  Deviations from 
spherical symmetry, or clumping of the dust will drive the effective radius up,
perhaps greatly, so this implied emission radius is really only a lower 
limit.  Since this limit is consistent with emission from either the
ejecta or the CSM, we are unable to use this to distinguish between the two 
possible sources for the emitting dust. 

Having the NIR continuum powered by the circumstellar interaction 
would help explain the similar temperatures seen in SN~1998S and SN~1997ab.  
Unlike the dust formation and IR echo scenarios which have energy sources that 
decline with time, the interaction would provide a fairly constant source of 
energy which could continuously heat the dust. Thus the observed temperature 
would not necessarily be expected to drop. The characteristic 1200~K
temperature
might be understood as the highest temperature allowed without destroying 
the dust grains.  Since the 1--2.4 \micron \ region is not sensitive to cooler
dust we would always expect to measure a $\approx$ 1200~K temperature as long
as the interaction is providing enough energy to keep a significant amount 
of dust at this temperature.

Thus, on balance we find that the luminous late-time NIR continuum in 
SN~1998S and SN~1997ab appears consistent with dust emission being powered by
a strong circumstellar interaction.  This interpretation 
is supported by the fact that of the previously observed supernovae 
exhibiting a late time infrared excess, five (SN~1994Y, SN~1993J, SN~1979C, 
SN~1980K, and SN~1985L) showed evidence of strong interaction with a 
surrounding circumstellar medium (\cite{Clocchiatti94}; \cite{CD94}; 
\cite{FLC96}; \cite{FM93}; \cite{FB90}; \cite{Fesen98}).  On the basis of our
data, we are unable to distinguish between direct shock heated dust and dust
heated by X-rays from the interaction zone.   X-ray observations of 
SN~1998S and SN~1997ab may be able to address this question.   

\section{Conclusions}

We have presented near-infrared spectra of SN~1998S from about three
months to almost a year after maximum.  At day 95, the spectrum was
quite similar to the day 192 spectrum of SN~1987A, exhibiting emission
from H, He, CO, and a variety of low ionization features. As this is
the third detection of CO in a very small sample of Type~II SNe with NIR 
spectra, CO emission may be a fairly common feature of these events.
Much more late-time NIR spectroscopy of Type~II SNe is needed to address
this question adequately.

An examination  of the CO emission in SN~1998S indicates that the
temperature of the CO region was $\gtrsim$ 2000 K at day 95, perhaps as
high as 3000--4000 K.  There appears to be some excess emission shortward of
2.28~\micron \ which might be a blue wing to the CO profile, which would 
indicate that the CO is moving at $\approx$ 2000 -- 3000 km~s$^{-1}$.  
The lack of strong bandhead structure indicates that the CO is probably moving 
at least as fast as was observed in SN~1987A.

The later spectra of SN~1998S exhibit multi-peak emission line profiles in the
hydrogen and helium lines. The high velocity peaks of the circumstellar
features
were shown to be emission from an equatorial ring, while the low velocity
peak is consistent with a shocked population of dense clumps.  This indicates 
that the circumstellar environment of SN~1998S has a strong disk shaped 
component.  

The post 200$^{\rm d}$ NIR spectrum of SN~1998S was dominated by a red 
continuum that is especially prominent rising to longer wavelengths in the data
of day 355.  This red continuum was also observed in the Type~IIn SN~1997ab 
more than 1064 days after maximum, and the striking similarity of the NIR 
spectra suggest a common cause for the continuum in both objects.  The emission
seen in these objects is likely related to near-infrared excesses seen in the
photometry of a number of other supernovae.  This continuum could be due to 
free-free emission, but only if the ejecta is optically thick and only if the 
density profile at the effective photosphere is rather shallow.  On the other 
hand, the NIR continuua in both SN~1998S and SN~1997ab are matched quite well 
by a 1200 $\pm 150$ K blackbody consistent with emission from hot dust.  

The inferred NIR luminosities are quite high 
($\gtrsim 10^{41}$ erg~s$^{-1}$ for SN~1997ab) and radioactive heating can be 
ruled out as a power source purely from energy requirement considerations.  
We find that the continuum is likely due to dust heated by the interaction 
with the CSM, either by direct shock-heating, or by absorption of X-rays from 
the interaction region.  While there is strong evidence that dust has formed
in the ejecta of SN~1998S, we cannot say whether the observed dust emission is
from the ejecta, or from pre-existing dust in the circumstellar medium.

\acknowledgments
We thank R. McCray, T. Geballe, R. Kirshner, and P. Garnavich for many helpful 
discussions, and the MDM staff for general observing support.  We would also 
like to thank W.P.S. Meikle for allowing us to use his SN~1987A data.
CLG's and RAF's research is supported by NSF Grant 98-76703.  PH and JCW's 
research is supported in part by NSF Grant 95-28110, NASA Grant NAG 5-2888, 
GO-8243 from the Space Telescope Science Institute with is operated by AURA, 
Inc., under NASA contract NAS 5-26555, NASA Grant LSTA-98-022 and a grant from 
the Texas Advanced Research Program.

\clearpage
\begin{deluxetable}{cccrr}
\tablecaption{Log of Near-Infrared Observations of SN 1998S \label{tab1}}
\tablehead{
        \colhead{Observation} &
        \colhead{Wavelength} &
        \colhead{Spectral} &
        \colhead{Spectroscopic} &
        \colhead{Infrared} \\
        \colhead{Epoch} &
        \colhead{Coverage} &
        \colhead{Resolution} &
        \colhead{Observation} &
        \colhead{Photometry} \\
        \colhead{(J.D. 2450895+)} &
        \colhead{(\micron)} &
        \colhead{(km s$^{-1}$)} &
        \colhead{Date} &
        \colhead{Date}
        }
\startdata
95      & 1.10 -- 1.35 & 425 & 22 Jun 1998 & 20 Jun 1998 \nl
        & 1.47 -- 1.80 & 345 & 20 Jun 1998 & 20 Jun 1998 \nl
        & 1.99 -- 2.40 & 400 & 20 Jun 1998 & 20 Jun 1998 \nl
\nl
225     & 0.95 -- 1.75 & 430 & 29 Oct 1998 & 01 Nov 1998 \nl
        & 1.99 -- 2.40 & 430 & 01 Nov 1998 & 01 Nov 1998 \nl
\nl
260     & 0.95 -- 1.75 & 430 & 04 Dec 1998 & 04 Dec 1998 \nl
        & 1.99 -- 2.40 & 430 & 04 Dec 1998 & 04 Dec 1998 \nl
\nl
355     & 0.95 -- 1.75 & 430 & 08--09 Mar 1999 & 08--09 Mar 1999 \nl
        & 1.99 -- 2.40 & 430 & 08--10 Mar 1999 & 08--09 Mar 1999 \nl
\enddata
\end{deluxetable}

\clearpage
\begin{deluxetable}{cccc}
\tablecaption{Near-Infrared Photometry of SN 1998S \label{tab2}}
\tablehead{
        \colhead{Epoch} &
        \colhead{\textit{J}} &
        \colhead{\textit{H}} &
        \colhead{\textit{K}} \\
        \colhead{(J.D. 2450895+)} &
        \colhead{(mag)} &
        \colhead{(mag)} &
        \colhead{(mag)}
        }
\startdata
95      & $14.50\pm0.10$ & $14.08\pm0.10$ & $13.40\pm0.10$ \nl
225     & $16.03\pm0.10$ & $14.65\pm0.10$ & $13.47\pm0.05$ \nl
260    & $16.38\pm0.10$ & $14.92\pm0.05$ & $13.47\pm0.05$ \nl
355     & $17.00\pm0.08$ & $15.27\pm0.03$ & $13.52\pm0.03$ \nl
\enddata
\end{deluxetable}

\clearpage
\begin{deluxetable}{clllc}
\tablecaption{Near-Infrared Line Identifications for SN~1998S \label{tab3}}
\tablehead{
        \colhead{Epoch} &
        \colhead{$\lambda_{\rm obs}$} &
        \colhead{Line}  &
        \colhead{$\lambda_{\rm lab}$} &
        \colhead{Observed} \\
        \colhead{(J.D. 2450895+)} &
        \colhead{(\micron)} &
        \colhead{Identification } &
        \colhead{(\micron)} &
        \colhead{Flux\tablenotemark{a}}
        }
\startdata
95      & 1.129 & \ion{O}{1}              &       1.1287 & 9 \nl
        & 1.194 & \ion{Mg}{1}             &       1.1828 & 6 \nl
        & 1.284 & \ion{H}{1} (Pa$\beta$)  &       1.2818 & 24 \nl
        & 1.503 & \ion{Mg}{1}             &       1.5031 & 4 \nl
        & 1.57 -- 1.79 & \ion{H}{1} (Br Series) & 1.570 -- 1.737  & 14 \nl
        &            & \ion{Si}{1}        &       1.5888 & " \nl
        &       & [\ion{Fe}{2}]           & 1.5994, 1.6440, 1.6772 & " \nl
        &       & [\ion{Si}{1}]           & 1.6068, 1.6454         & " \nl
        &       & \ion{Mg}{1}             &       1.7109 &  " \nl
        & 2.062 & \ion{He}{1}             &       2.0581 & 1 \nl
        & 2.171 & \ion{H}{1} (Br$\gamma$) &       2.1655 & 6 \nl
        & 2.29 -- 2.43 & CO ($\Delta v=2$)  & 2.294 -- 2.512 & 19 \nl
225     & 1.085 & \ion{He}{1}, \ion{H}{1} (Pa$\gamma$)  & 1.0830, 1.0938 & 13
\nl
        & 1.13  & \ion{O}{1}              &       1.1287 & 3 \nl
        & 1.12  & \ion{Mg}{1}             &       1.1828 & 2 \nl
        & 1.285 & \ion{H}{1} (Pa$\beta$)  &       1.2818 & \nl
        & 2.29 -- 2.40 & CO ($\Delta v=2$)  & 2.294 -- 2.512& 5 \nl
260     & 1.079 & \ion{He}{1}, \ion{H}{1} (Pa$\gamma$)  & 1.0830, 1.0938 & 9
\nl
        & 1.281 & \ion{H}{1} (Pa$\beta$)  &       1.2818 & 5 \nl
355     &1.076 & \ion{He}{1}, \ion{H}{1} (Pa$\gamma$)  & 1.0830, 1.0938 & 6 \nl
        & 1.277 & \ion{H}{1} (Pa$\beta$)  &       1.2818 & 2 \nl
\enddata
\tablenotetext{a}{in units of $10^{-14}$ erg s$^{-1}$ cm$^{-2}$ }
\end{deluxetable}
\clearpage
\appendix
\section{Carbon Monoxide Emission Models}
In order to analyze the SN~1998S CO spectrum, a total of 96 parameterized,
spherically symmetric, and highly simplified models were calculated.  In all of
our models, the CO bands are optically thick and, consequently, cannot be
treated in the Sobolev approximation.  Instead, we solve the non-relativistic
radiation transport equations in comoving frame using the method of Mihalas,
Kunatz \& Hummer (1975, 1976).

Power law density profiles have been studied with $\rho \propto r^{-n}$ with
$n=4 ... 9$. We assumed temperature profiles given by the analytic solution
for scattering dominated atmospheres (\cite{chand45}) which represents
a good first order solution for  scattering dominated photospheres
(e.g. \cite{hof91}).  Thus, we use the following expression:
$$ T = 0.9306 ~T_{eff} \left[ \int^{\tau_{ross}}_0 \left( R^2 \over r^2
\right) d \tau_{ross} + C \right] ^{0.25}$$
where
$$R \equiv r(\tau_{ross}=1), $$
and the Rosseland optical depth $\tau_{ross} $ is defined by the
corresponding  opacity  $$ \chi_{ross} \equiv {4 \sigma T^3 \over \pi} / \int
{1 \over  \chi_\nu} {\partial B_\nu \over \partial T} d \nu. $$
The parameter C represents the geometrical dilution of the radiation field in
the optically thin case and it is taken as
$$C = {\mu_0 \over 3} + {1 - \mu_0 \over \sqrt 3 } $$ with
$$\mu_0 (r)= \sqrt{max\biggl\lbrack0.,1 - \biggl({R \over r}\biggr)^2
\biggr\rbrack}. $$

Effective temperatures $T_{eff}$ and photopsheric radii $R_{star}$ have been
studied in the range between 3000 to  5500 K and $10^{14 ... 15}$ cm,
respectively. For low temperatures, the photosphere is ionization bound. The
matter above the ionization region contributes little to the continuum optical
depth and, consequently, the mass of the envelope above the photosphere is a
free parameter which has been varied between 0.2 and 5 $M_\odot$. Note that
changing the mass above the photophere has a very similar effect to a variation
of the C O abundances.  Solar abundances were assumed.  Bound-free and
free-free opacities and the occupation numbers have been included in local
thermodynamical equilibrium which has been found to be a good approximation for
the IR (within our model assumptions).  The full time dependent rate equations
have been solved for the formation of CO including radiative, collisional and
charge exchange processes according to \cite{petu89} with
modifications according to \cite{arvett96}.  The cross sections have been taken
from \cite{petu89} and \cite{dal90}, and references therein.

For the formation of CO, we include three-body association
$$  C + O + He \rightarrow CO + He,$$
$$  C^+ + O + He\rightarrow CO^+ + He, $$
$$  C + O^+ + He \rightarrow CO^+ + He,$$
and radiative associations
$$ C + O \rightarrow CO + \gamma,$$
$$ C^+ + O \rightarrow CO^+ + \gamma $$
$$ C + O^+ \rightarrow CO^+ + \gamma. $$
For the dissociation, we include
$$ CO^+ + e\rightarrow C + O,$$
$$ He + CO\rightarrow C + O + He,$$
$$ He^+ + CO\rightarrow C^+ + O + He,$$
charge exchange reactions,
$$ CO^+ + O \rightarrow CO  + O^+$$
and radiative dissociation
$$ CO  + \gamma\rightarrow C + O^+,$$
$$ CO^+ + \gamma\rightarrow C + O^+,$$
$$ CO + \gamma \rightarrow CO^+ +e $$
with  photon energies above  11.1, 8.3 and  14 eV, respectively.
For the UV flux, we assume a geometrically diluted black body with $T=T_{eff}$.

The time-dependent rate equations were then integrated for a given time.  For
simplification, the photospheric radius and luminosity is assumed to be
constant over this period of time.  In Figure 12, we give the change of the CO
abundance as a function of the assumed integration/formation time normalized
to a duration of 90 days for the formation.  In this example, the first
overtone of CO is formed between 1.1 and 1.5 $\times 10^{15}$ cm.
The CO-abundance may be uncertain by a factor of 70\% due to the uncertainty
in the duration CO formation in SN1998S. Despite this uncertainty,
the decoupling regions of photons change by only about 1/2 of a scale
height, which implies a small radial change.  Therefore, the results on the
final spectra are rather insensitive. In our example, the CO emission changes
by about 15\%.  For all the models in Figure 7, the duration of the
CO-formation was set to 30 days.

\clearpage
\begin{figure}
\figurenum{1}
\plotone{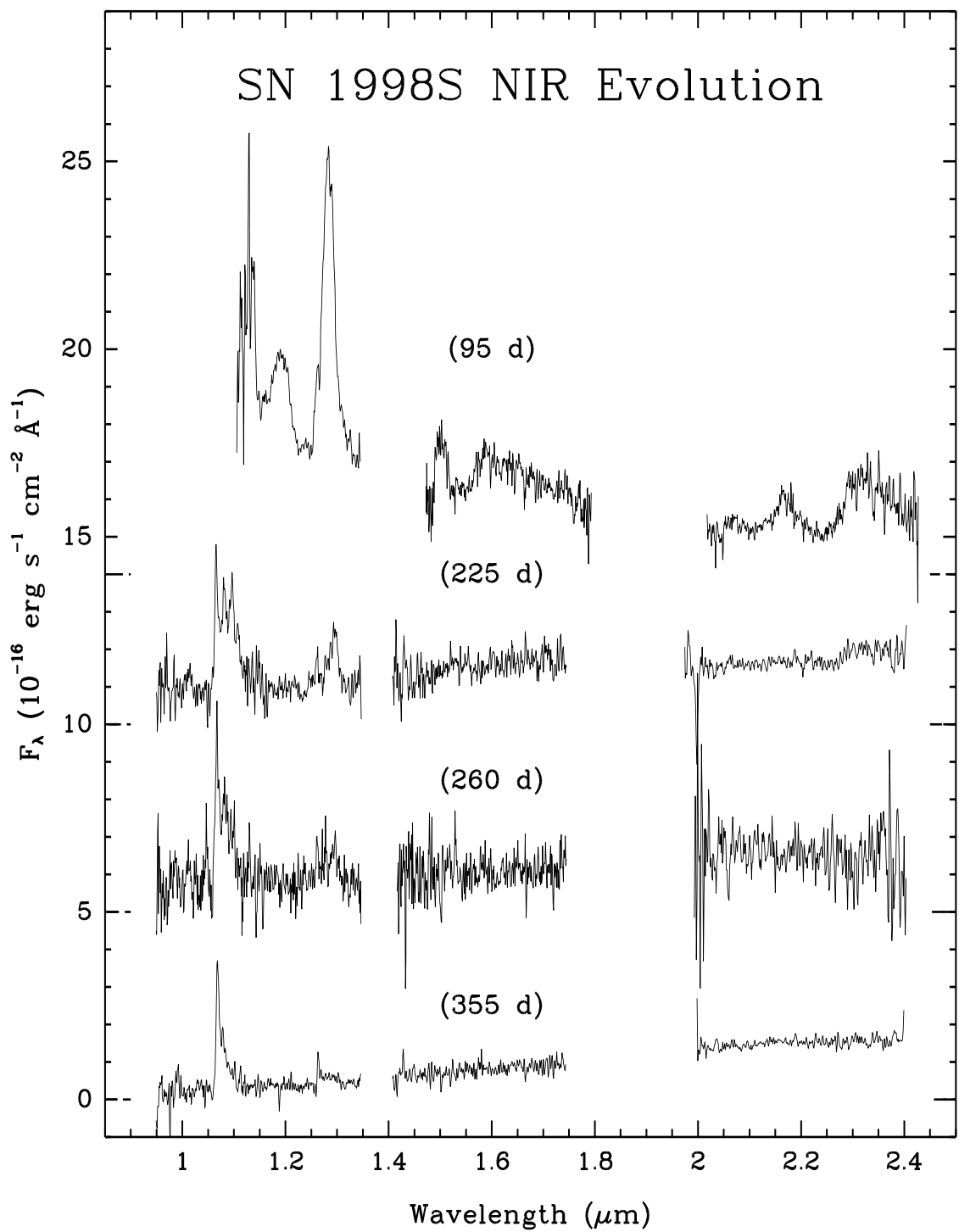}
\caption{The time evolution of the NIR spectrum of
SN~1998S.  The data have been shifted vertically for clarity, and the
zero-points are marked along the edges. The data are presented in vacuum
wavelengths and are shifted to the rest frame of the host galaxy, NGC~3877.}
\end{figure}

\clearpage
\begin{figure}
\figurenum{2}
\plotone{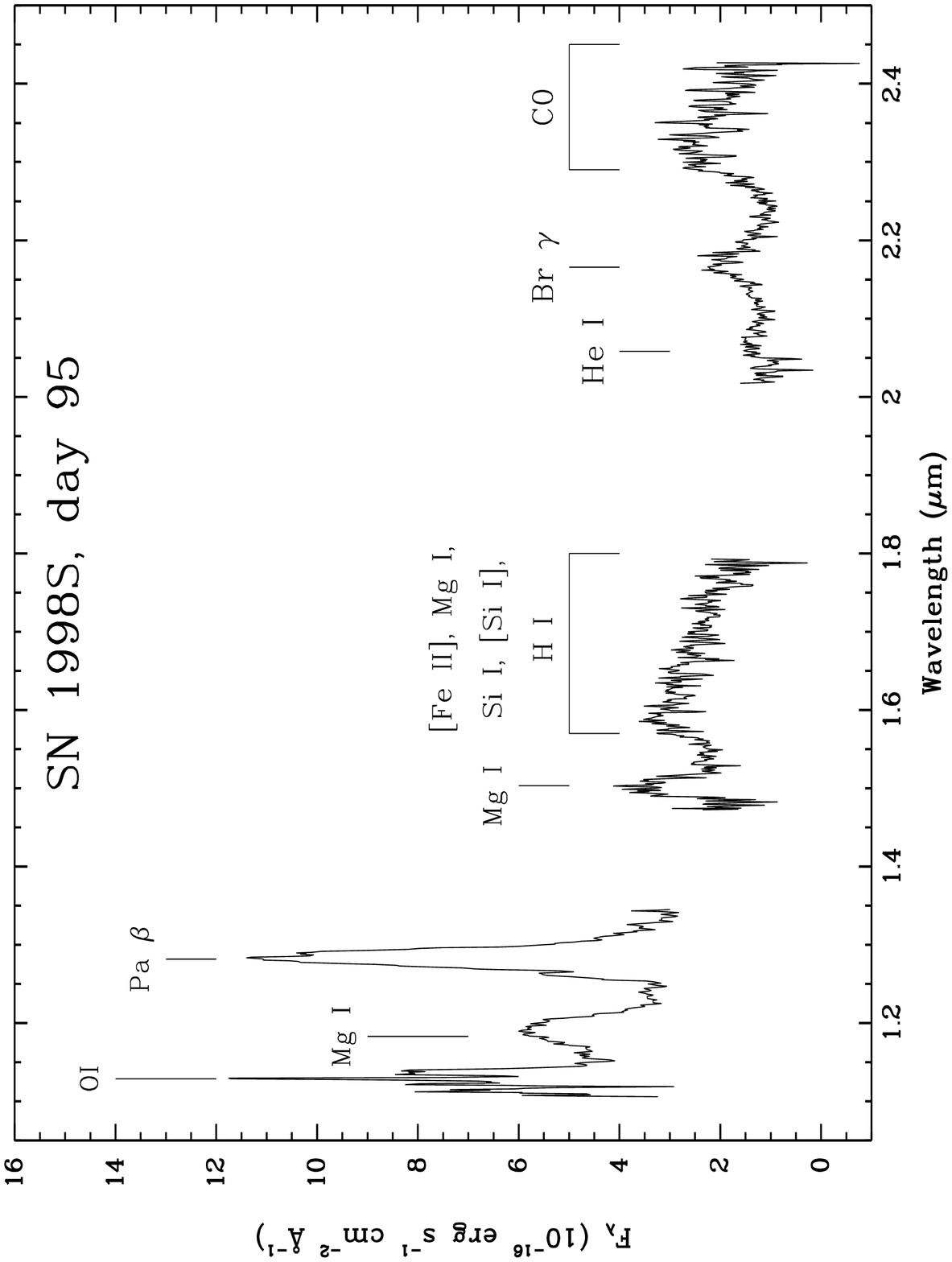}
\caption{The day 95 NIR spectrum of SN~1998S. The data are
presented in vacuum wavelengths and are shifted to the rest frame of the
host galaxy, NGC~3877.}
\end{figure}

\clearpage
\begin{figure}
\figurenum{3}
\plotone{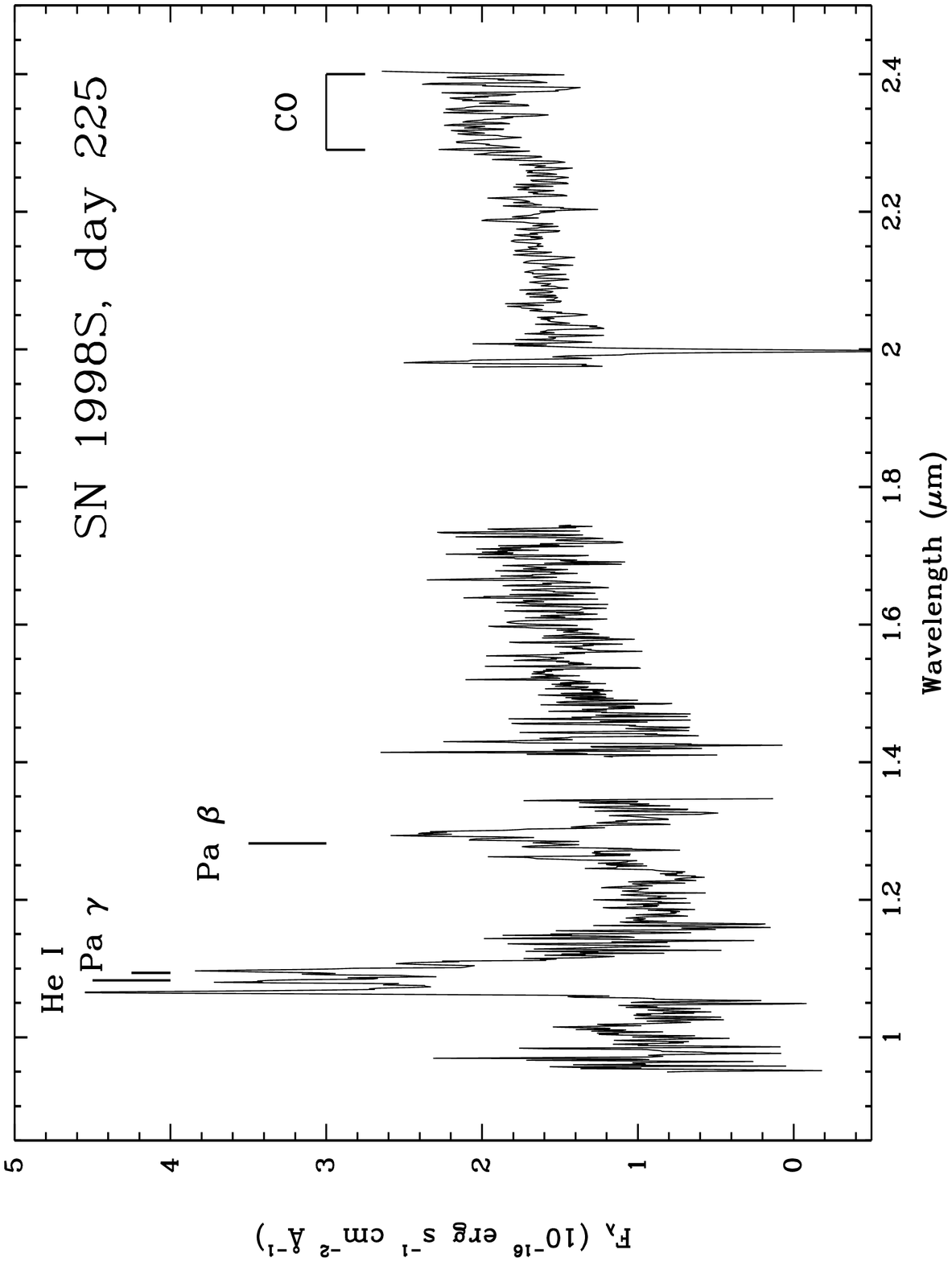}
\caption{The day 225 NIR spectrum of SN~1998S.  The data
are presented in vacuum wavelengths and are shifted to the rest frame of the
host galaxy, NGC~3877.}
\end{figure}

\clearpage
\begin{figure}
\figurenum{4}
\plotone{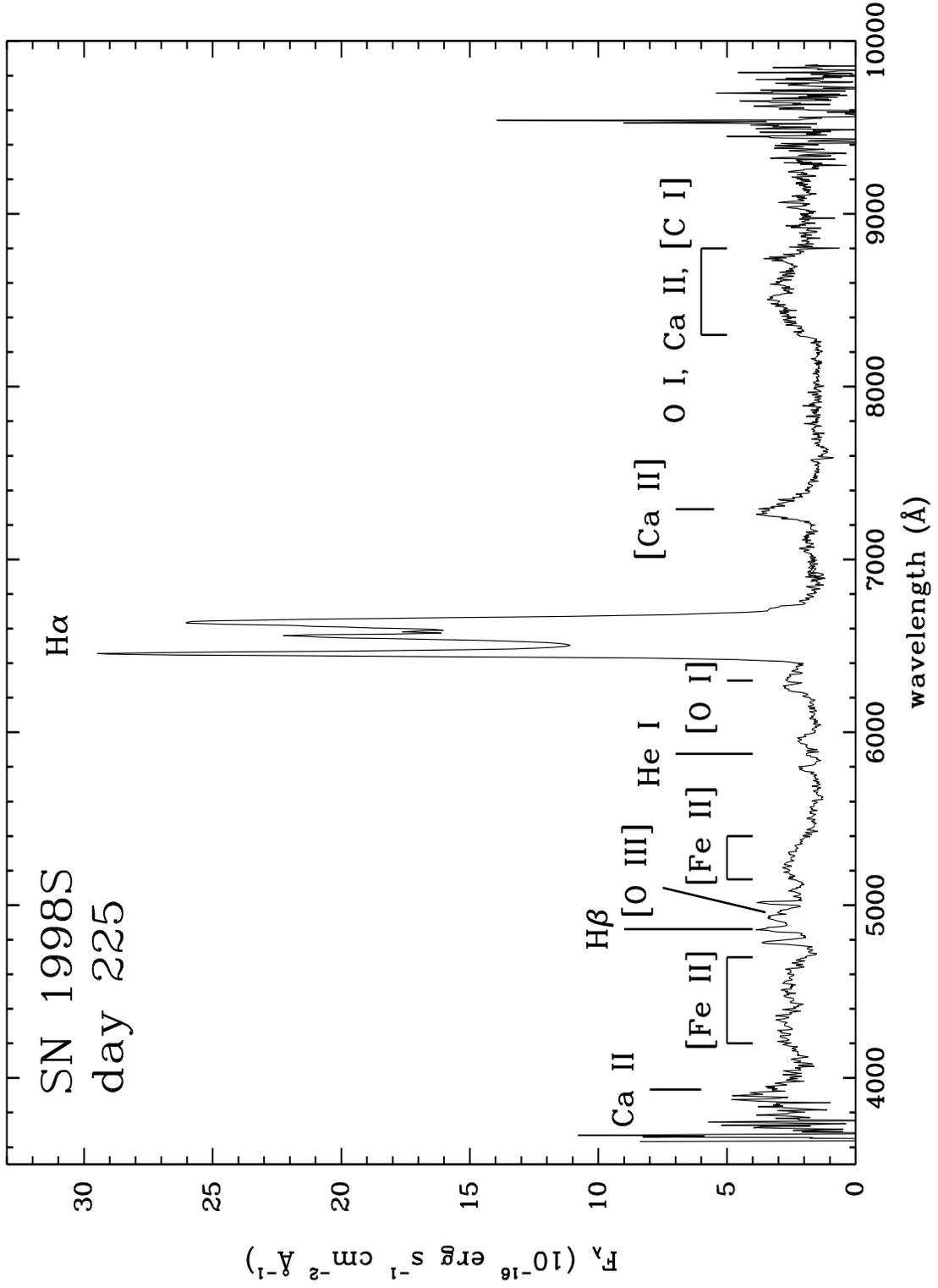}
\caption{The day 225 optical spectrum of SN~1998S. The data are shifted to the 
rest frame of the host galaxy, NGC~3877.}
\end{figure}

\clearpage
\begin{figure}
\figurenum{5}
\plotone{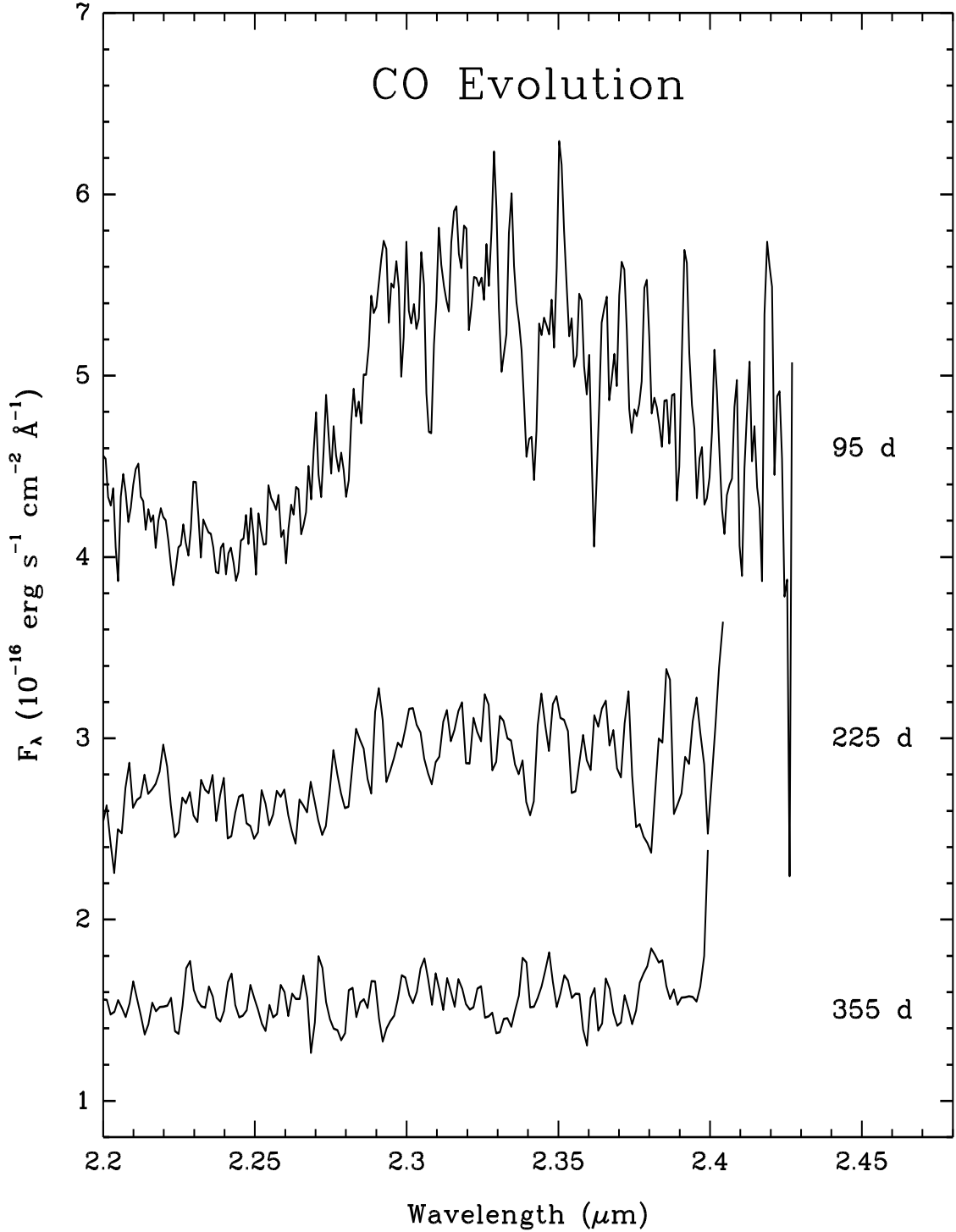}
\caption{Portions of the \textit{K}-band spectrum of
SN~1998S for t=95$^{\rm d}$, 225$^{\rm d}$, and 355$^{\rm d}$, showing the time
evolution of the CO feature.  For clarity, the 95$^{\rm d}$ and 225$^{\rm d}$
spectra have been shifted vertically by three and one units, respectively. The
data are presented in vacuum wavelengths and are shifted to the rest frame of
the host galaxy, NGC~3877.}
\end{figure}

\clearpage
\begin{figure}
\figurenum{6}
\plotone{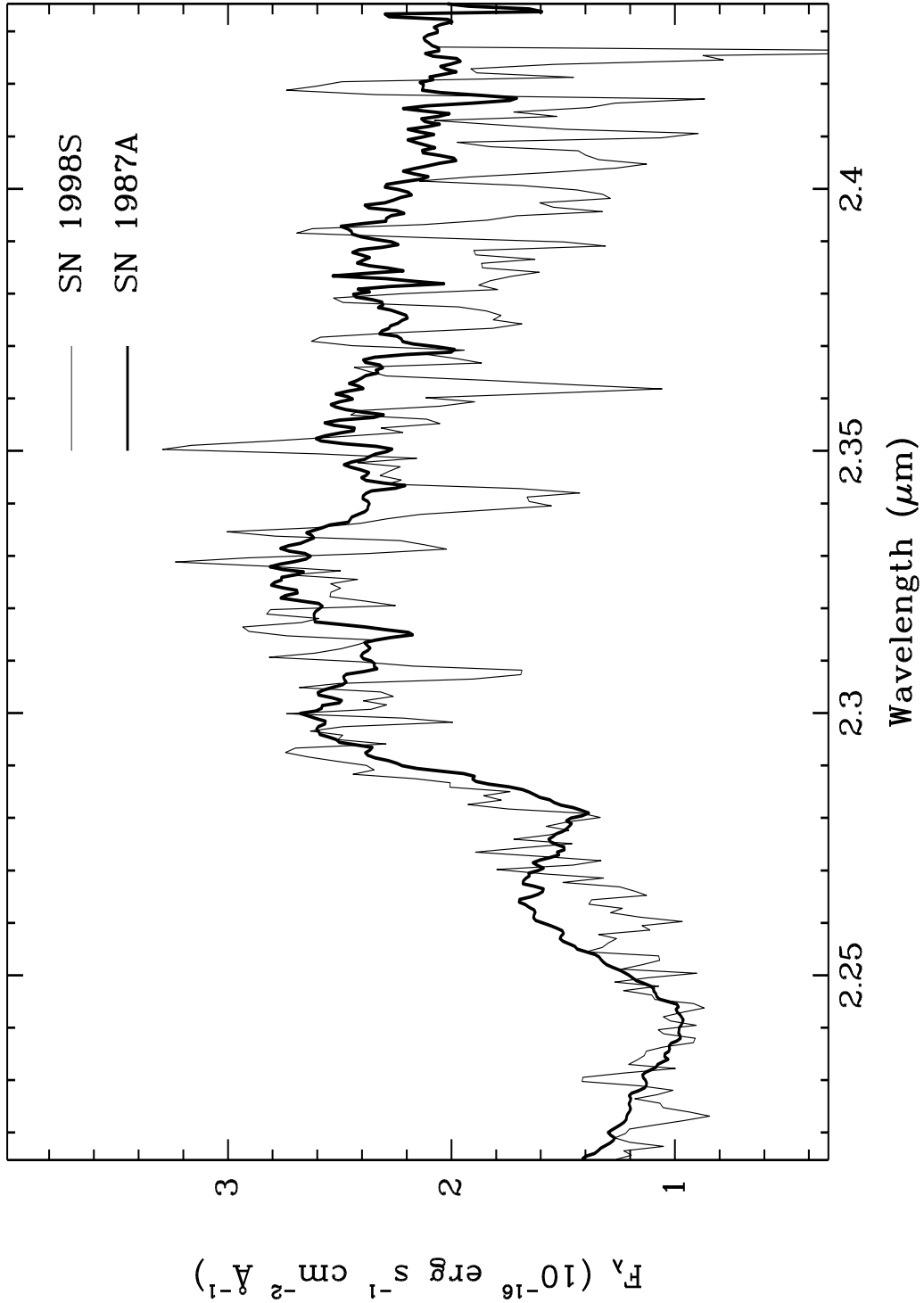}
\caption{The day 95 SN~1998S spectrum and the day 192
spectrum of SN~1987A presented by M89.  To facilitate the comparison, we have
shifted both spectra to the rest frame of the supernova, and we have scaled the
SN~1987A data to match the SN~1998S data, at the continuum near 2.24 \micron \
and near the CO peak in the 2.30 -- 2.32 \micron \ region.}
\end{figure}

\clearpage
\begin{figure}
\figurenum{7}
\plotone{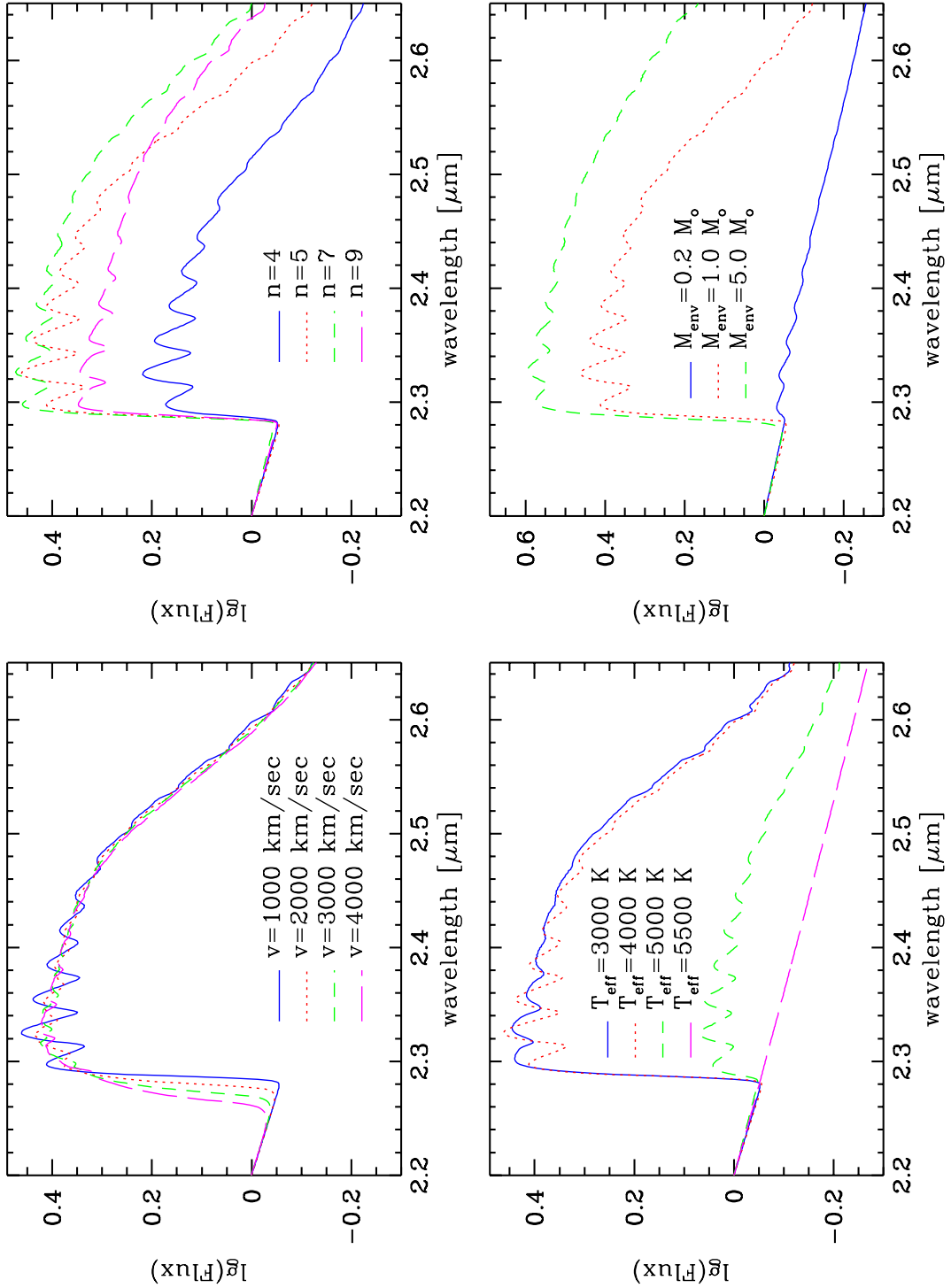}
\caption{Dependence of the flux in the first overtone of
CO on the free model parameters: velocity, density slope, temperature, and mass
of the envelope above the photosphere. CO formation was assumed to set in 30
days earlier (see text and Appendix A).}
\end{figure}

\clearpage
\begin{figure}
\figurenum{8}
\plotone{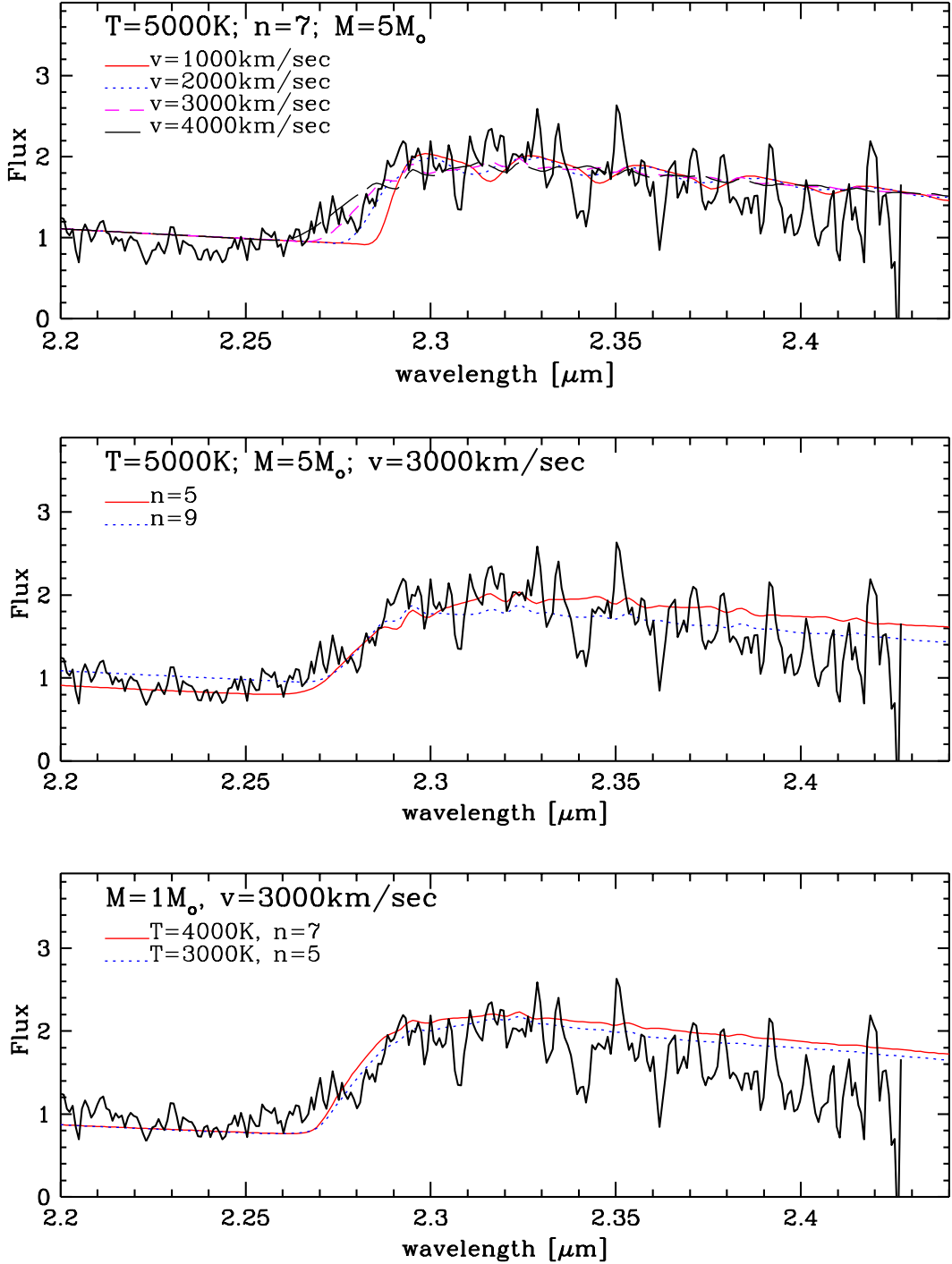}
\caption{Comparison of the day 95 CO emission in SN~1998S
with models for various expansion velocities, temperatures, density
profiles, and envelope mass.}
\end{figure}

\clearpage
\begin{figure}
\figurenum{9}
\plotone{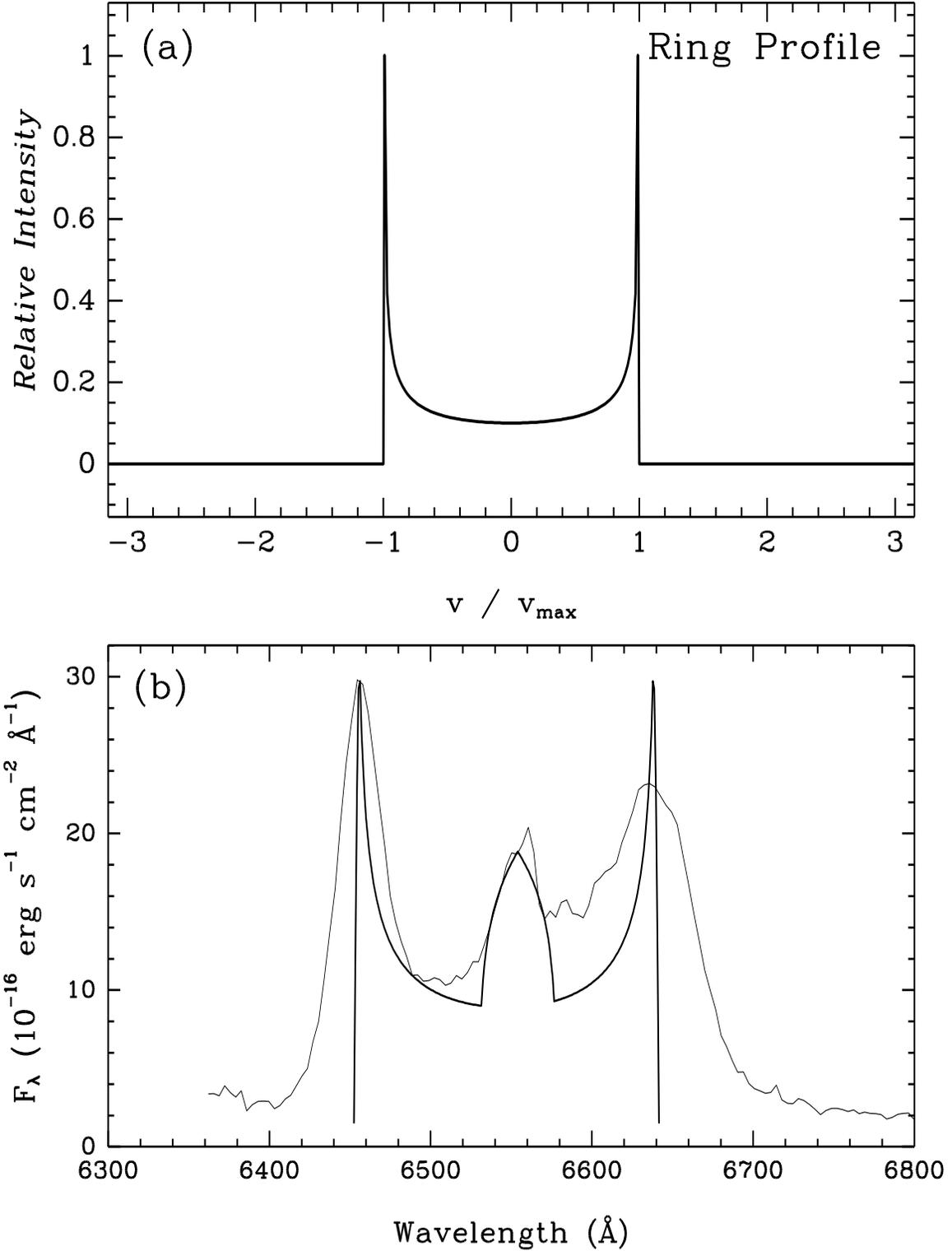}
\caption{(a) The theoretical velocity profile for line
emission from a thin, uniformly emitting ring of gas expanding at velocity
$V_{Max}$. (b) The observed H$\alpha$ profile at day 225, and a line profile
from a two component ``toy model'' for this emission.  The outer regions of the
profile are emission from a thin annulus expanding at 4300 km~s$^{-1}$.  The
central peak is emission from a population of dense clouds with a number
density $n \propto \rho_{c}^{-1.9}$, and a maximum cutoff velocity
$u_{c_{max}} = 1000$ km~s$^{-1}$.  See text for details.}
\end{figure}

\clearpage
\begin{figure}
\figurenum{10}
\plotone{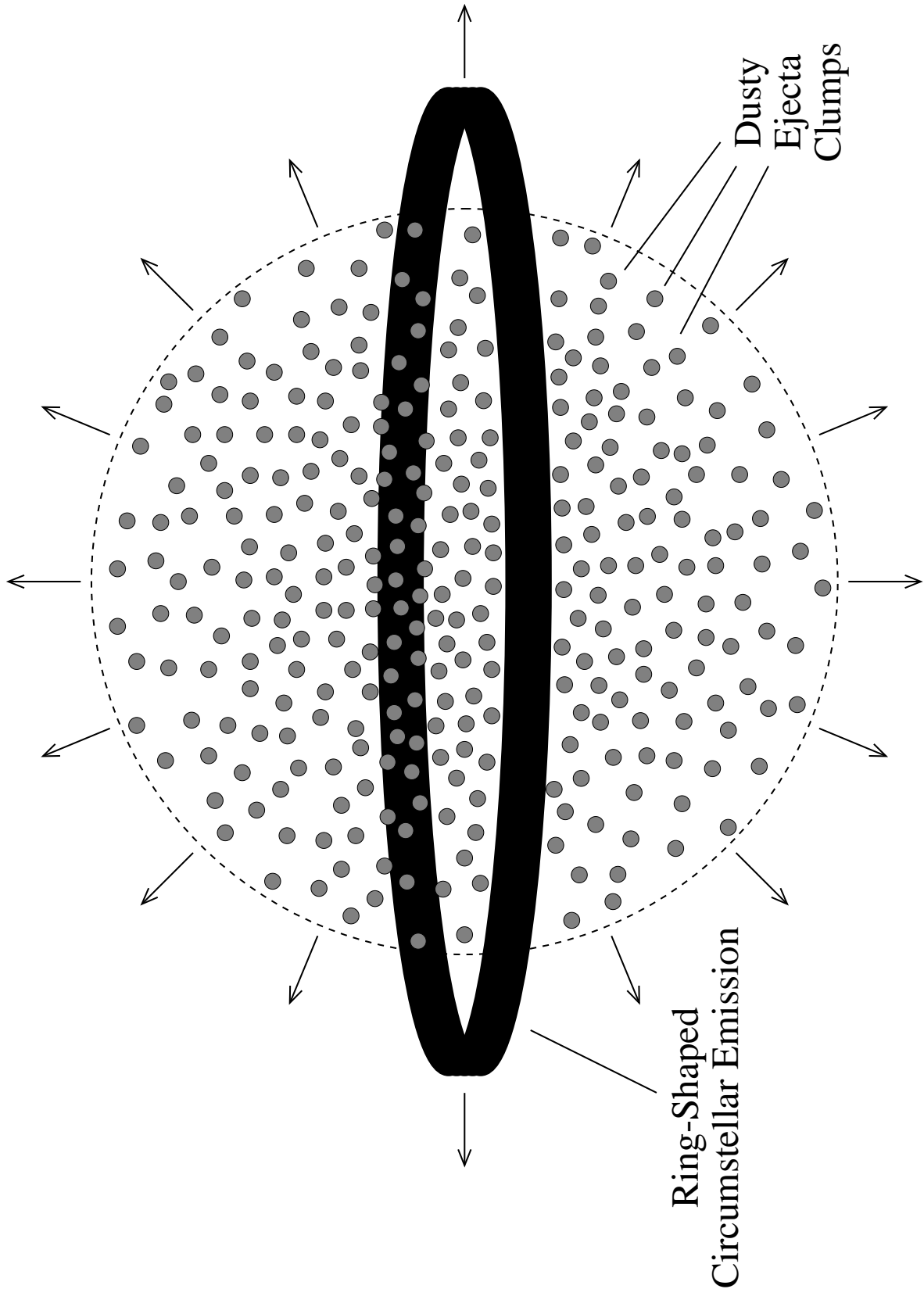}
\caption{Schematic representation of the geometry proposed
for SN~1998S.  The circumstellar emission comes from an equatorial ring where
the blast wave is running into a dense circumstellar disk.  The far side of the
circumstellar ring is obscured by dust which has formed in dense clumps in the
ejecta.}
\end{figure}

\clearpage
\begin{figure}
\figurenum{11}
\plotone{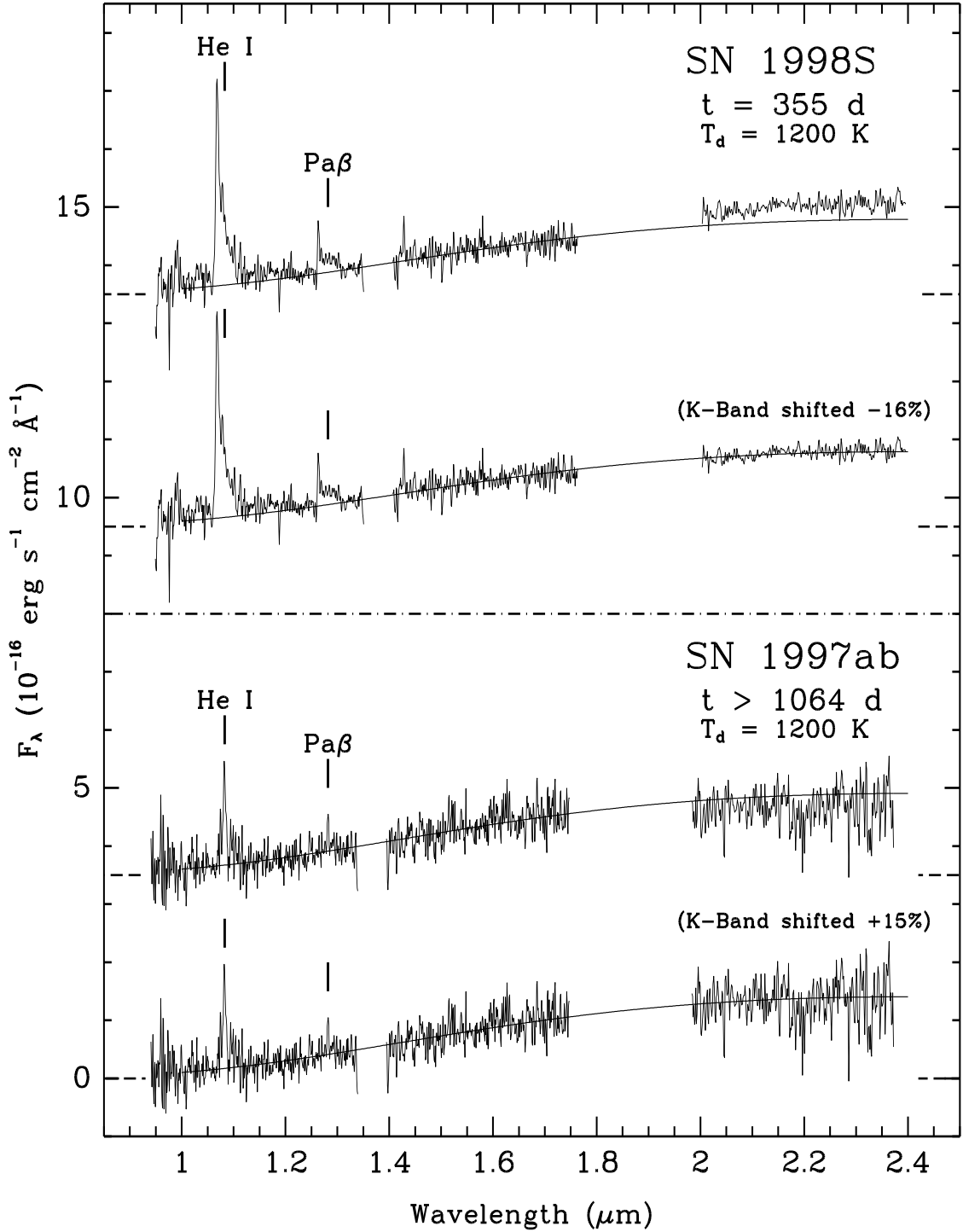}
\caption{Day 355 NIR spectrum of SN~1998S and day 1064+
for the Type IIn object SN~1997ab. Overplotted are blackbody curves for
1200 K.  The flux level of the blackbody curves have been matched to the
flux level of the \textit{H}-band region, and in the 2nd and 4th spectra, the
\textit{K}-band fluxes have been shifted by the amounts labeled to better
match the blackbody curves.}
\end{figure}

\clearpage
\begin{figure}
\figurenum{12}
\plotone{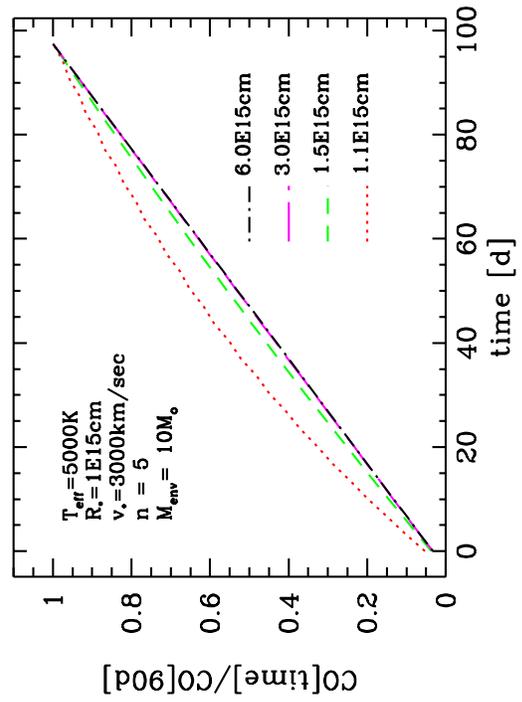}
\caption{CO formation as a function of time clearly shows
the dependence of the CO abundance on the duration of the formation.  The
spectral features of the first overtone of CO are formed between 1.1 and 1.5
$\times 10^{15}$ cm.}
\end{figure}

\end{document}